\documentclass[apj]{emulateapj}
\usepackage{apjfonts}
\usepackage{epsfig}
\usepackage{amsmath}
\usepackage{amssymb}
\usepackage{amsfonts}
\usepackage{natbib}          
\shorttitle{X-RAY AGN IN ABELL 85 \& ABELL 754}
\shortauthors{SIVAKOFF ET AL.}

\newcommand\mion[2]{\textrm{\ion{#1}{#2}}}

\defcitealias{MKK+2006}{M06}
\defcitealias{MMK2007}{M07}
\defcitealias{CZ2003}{CZ03}

\slugcomment{Accepted for publication in ApJ}

\begin{document}

\title{Wide-Field Chandra X-Ray Observations of AGN in Abell 85 \& Abell 754}

\author{
Gregory R. Sivakoff\altaffilmark{1},
Paul Martini\altaffilmark{1},
Ann I. Zabludoff\altaffilmark{2},
Daniel D. Kelson\altaffilmark{3},
John S. Mulchaey\altaffilmark{3}
}

\altaffiltext{1}{
Department of Astronomy,
The Ohio State University,
4055 McPherson Laboratory
140 W. 18th Avenue, Columbus, OH 43210-1173, USA;
sivakoff@astronomy.ohio-state.edu
}
\altaffiltext{2}{
Steward Observatory,
University of Arizona,
933 N Cherry Ave., Rm. N204
Tucson, AZ 85721-0065, USA
}
\altaffiltext{3}{
Carnegie Observatories,
813 Santa Barbara St.,
Pasadena, CA 91101-1292, USA
}

\begin{abstract}
To better understand the mechanism or mechanisms that lead to AGN activity
today, we measure the X-ray AGN fraction in a new sample of nearby clusters and
examine how it varies with galaxy properties, projected cluster-centric
radius, and cluster velocity dispersion. We present new wide-field {\it
Chandra X-ray Observatory} observations of Abell 85, Abell 754 and the
background cluster Abell 89B out to their virial radii. Out of seventeen X-ray
sources associated with galaxies in these clusters, we classify seven as X-ray
AGN with $L_{X,B} > 10^{41} {\rm \, erg \, s}^{-1}$. Only two of these would be
classified as AGN based on their optical spectra. We combine these observations
with archival data to create a sample of X-ray AGN from six $z < 0.08$ clusters
and find that $3.4^{+1.1}_{-0.8}\%$ of $M_R < -20$ galaxies host X-ray AGN with
$L_{X,B} > 10^{41} {\rm \, erg \, s}^{-1}$. We find that more X-ray AGN are
detected in more luminous galaxies and attribute this to larger spheriods in
more luminous galaxies and increased sensitivity to lower Eddington-rate
accretion from black holes in those spheroids. At a given X-ray luminosity
limit, more massive black holes can be accreting less efficiently, yet still be
detected. If interactions between galaxies are the principal drivers of AGN
activity, then the AGN fraction should be higher in lower velocity dispersion
clusters and the outskirts of clusters. However, the tendency of the most
massive and early-type galaxies to lie in the centers of the richest clusters
could dilute such trends. While we find no variation in the AGN fraction with
projected cluster-centric radius, we do find that the AGN fraction increases
significantly from $2.6^{+1.0}_{-0.8}\%$ in rich clusters to
$10.0^{+6.2}_{-4.3}\%$ in those with lower velocity dispersions.
\end{abstract}

\section{Introduction}

\setcounter{footnote}{0}

What is the principal driver of Active Galactic Nuclei (AGN) in the nearby
Universe? Major mergers between gas-rich galaxies are largely accepted as the
dominant fueling mechanism \citep[e.g.,][]{BH1992} for the luminous quasar
population. Galaxy harassment, where dynamical instabilities driven by
high-speed fly-by interactions efficiently channel gas to the centers of
galaxies, has also been invoked as a mechanism for fueling AGN \citep{MKL+1996,
LKM1998}. In both scenarios, higher AGN fractions are predicted for environments
where gas-rich galaxies are likely to interact with one another. Although galaxy
densities are high, such interactions are not favored in the centers of rich
clusters, whose galaxies are less (cold) gas-rich than their counterparts in the
field \citep[e.g.,][]{GH1985} and where the large relative velocities betweens
galaxies inhibits actual mergers. Higher fractions of AGN are expected for lower
velocity dispersion structures. The AGN fraction at the outskirts of clusters
should also be larger as a higher fraction of gas-rich galaxies are found toward
the outskirts of clusters and infalling structures with lower velocity
dispersions may not yet have virialized. Some of this picture has been supported
by numerous studies of clusters that identified AGN by their optical spectra
observed a substantial decrease in the number of cluster AGN relative to the
field \citep{G1978}. Specifically, \citet{DTS1985} measured a decrease from
$5\%$ to $1\%$ in AGN residing in bright galaxies.

On the other hand, a large fraction of elliptical galaxies
($\sim 35$--$45\%$) contain low-ionization nuclear emission-regions
\citep[LINERs;][]{HFS1997c}, many of which may be ionized by the accretion
disk of a low-luminosity AGN \citep{HFS1993}. These elliptical galaxies comprise
a higher fraction of the galaxy populations in the high surface density regions
at the centers of clusters \citep{D1980}, and in particular the most luminous
elliptical galaxies ($M_R < -22$) that are much more centrally concentrated
\citep{TK2006}. Toward the outskirts of clusters, progressively higher fractions
of poststarburst and starforming galaxies are found
\citep[e.g.,][]{D1980,FFF+1998}. Therefore, a relation between AGN and
early-type galaxies could dilute or even reverse the trends predicted by
gas-rich mergers or galaxy harassment.

To gain leverage on these issues, it is critical to improve on the range of
environments probed by past studies. We are continuing a program that measures
the AGN fraction with environment, probing cluster environments for these
indirect signatures of AGN fueling mechanisms. To identify the AGN we use X-ray
observations. Galaxy studies in the nearby Universe
\citep[e.g.,][]{GGS2003,KF2004,SJF+2007} indicate that contributions from the
other potential sources of luminous X-ray emission besides an AGN, namely X-ray
binaries and the hot interstellar medium (ISM), only exceed X-ray luminosities
of $\sim 10^{41} {\rm \, erg \, s}^{-1}$ for the most massive or massively
star-forming galaxies. Thus, AGN can be identified down to relatively low X-ray
luminosities by considering galactic parameters such as their optical luminosity
and star formation rate. In addition, X-ray observations can identify AGN that
lack obvious spectral signatures in visible wavelength spectra. Such signatures
could potentially be absent due to selection effects \citep[e.g., optical
dilution of low-luminosity AGN,][]{MKM+2002,MFC2002}, obscuration
\citep[e.g.,][]{M2002}, or different accretion modes \citep[e.g., radiatively
inefficient accretion flows that do not produce emission lines,][]{YN2004}.

In the most detailed study of
X-ray AGN in clusters to date, \citet[][hereafter M06]{MKK+2006} studied eight
low-redshift ($0.06 < z < 0.31$) galaxy clusters and found that $\sim 5\%$ of
bright ($M_R < -20$) cluster galaxies contain AGN with $L_{X,B} > 10^{41} {\rm
\, erg \, s^{-1}}$, where $L_{X,B}$ is the broad ($0.3$--$8 {\rm \, keV}$) band
X-ray luminosity. Most of these X-ray identified AGN lacked obvious AGN spectral
signatures in visible wavelength spectra. In this sample, the $L_{X,B} > 10^{42}
{\rm \, erg \, s^{-1}}$ X-ray AGN were centrally concentrated \citep[hereafter
M07]{MMK2007}. When fainter X-ray AGN with $L_{X,B} > 10^{41} {\rm
\, erg \, s^{-1}}$ were included, no central concentration was found, although
they had limited sensitivity to radial variations in their more distant
clusters. While the increase in AGN fraction of bright X-ray AGN is more
consistent with the increased concentration of bright elliptical galaxies
outlined above as opposed to predictions from galaxy interactions, it is
somewhat surprising that the lower luminosity AGN are not also centrally
concentrated under such a model.
 
While X-ray AGN fractions in nearby clusters have been previously measured, past
observations have concentrated on the cores of clusters. In \citet[hereafter
M07]{MMK2007}, 90\% of the galaxies were within $0.5 \, r_{200}$, where
$r_{200}$ is the physical radius within which the mean density of a virialized
cluster exceeds the critical density at that redshift by a factor of 200. The
outskirts of clusters, and their different environment, are relatively
unexplored. This highlights the value for X-ray observations that can identify
X-ray AGN beyond the cores of clusters. Nearby clusters allow the most sensitive
measurements at both visible and X-ray wavelengths. But such observations must
be made over wide fields-of-view (FOVs) to cover the entire cluster. With its
superb spatial resolution, the {\it Chandra X-ray Observatory} is ideal for
detecting a central AGN; however, its widest FOV (using the ACIS-I detectors) is
only $\sim 17\arcmin \times 17
\arcmin$. For nearby clusters, this does not provide adequate coverage out to
$r_{200}$.

To attain the best measurements on the radial distribution of AGN for comparison
to the opposing predictions, we undertook wide-field {\it Chandra} observations
of two $z \sim 0.06$ clusters, Abell 85 and Abell 754. In X-rays, both Abell 85
\citep{KSR2002,DLF2005} and Abell 754 \citep[e.g.,][]{MMV+2003} show evidence of
recent mergers of multiple components; both clusters show evidence of cold
fronts in their intracluster medium (ICM). In particular, Abell 754 is often
used as a prototype of a major cluster - cluster merger, with the peak of its
X-ray emission well offset from the major galaxy clumps identified by optical
data \citep{ZZ1995}, while there is no such offset in Abell 85, where smaller
structures appear to be falling on to the major component of Abell 85
\citep{DFG+1998}. Both clusters already have detailed optical spectroscopy
\citep[hereafter CZ03]{CZ2003} that established cluster membership and
measured other spectral properties.
We present the analysis of these observations in \S~\ref{sec:obs}. We add these
clusters and Abell 89B, an additional cluster in the Abell 85 FOV, to three
clusters from the \citetalias{MKK+2006} study to form a sample of $z \lesssim
0.08$ clusters in \S~\ref{sec:sample}.
In \S~\ref{sec:galaxy}, we detail the
identification of sources as X-ray AGN and spectroscopically identified AGN, and
compare their properties (photometric and radial distribution) to the
underlying cluster population. We present the dependence of AGN fraction on
velocity dispersion and redshift in \S~\ref{sec:cluster}.
Finally, we discuss our conclusions
in \S~\ref{sec:end}. 
All errors presented indicate
the double-sided $1\sigma$ confidence interval%
\footnote{We note that previous error bars on the AGN fraction presented
single-sided $90\%$ confidence intervals, which are slightly larger
\citepalias{MKK+2006,MMK2007}.}.
Throughout this paper we assume that the cosmological parameters are
$(\Omega_{\rm M}, \Omega_{\rm \Lambda}, h) = (0.3, 0.7, 0.7)$, where $H_0 = 100
\, h {\rm \, km \, s}^{-1} {\rm \, Mpc}^{-1}$. All absolute magnitudes and
luminosities are presented in their rest-frame.

\section{{\it Chandra} Observations}
\label{sec:obs}

\begin{figure*}
\plotone{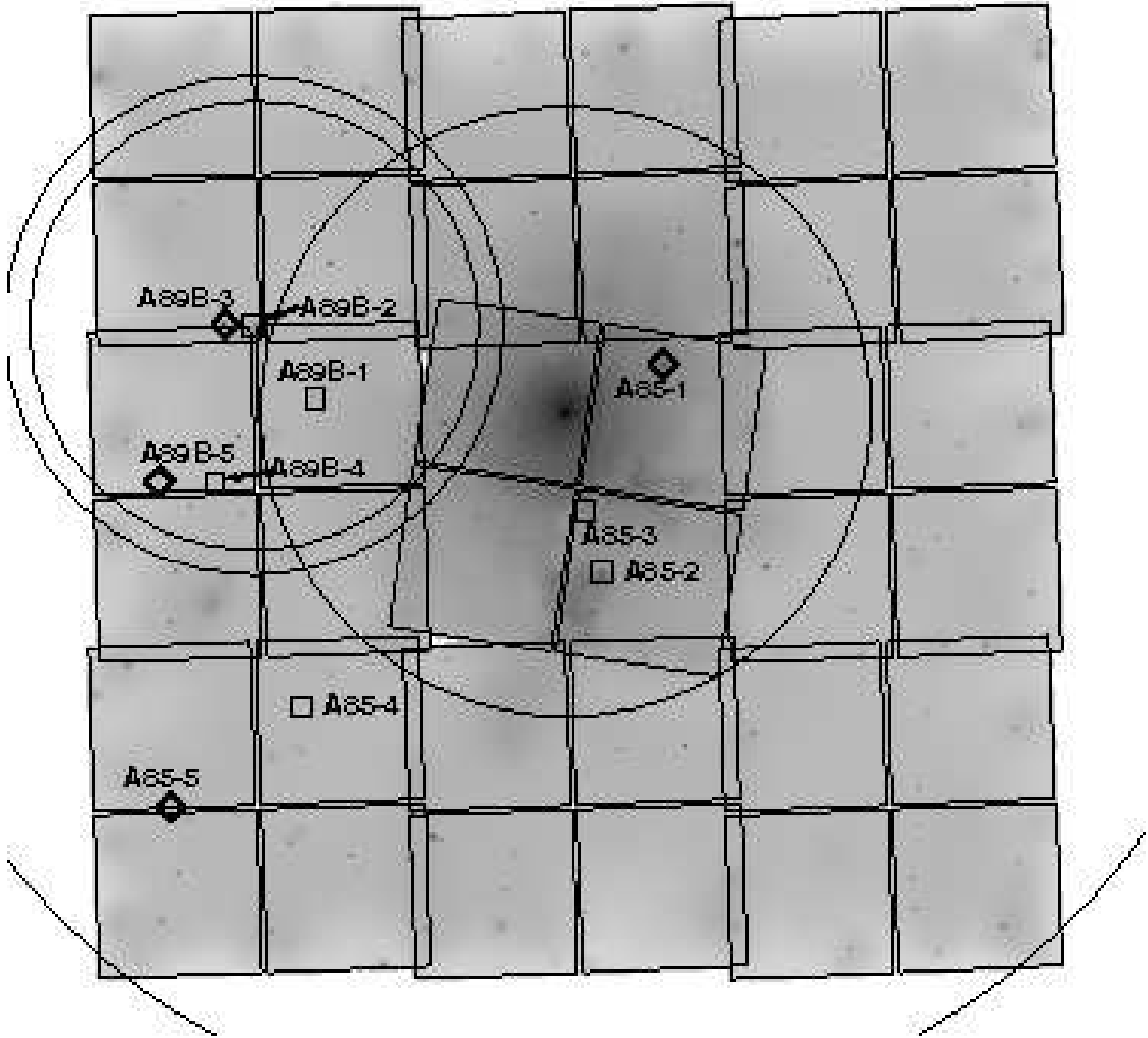}
\caption{
Adaptively smoothed {\it Chandra} mosaic of Abell 85 with individual {\it
Chandra} FOVs indicated. An arcsinh scaling has been applied to bring out both
point sources and intracluster gas. Inner and outer circles are used to display
the 1 Mpc radius and $r_{200}$, respectively, for both the Abell 85 (center) and
Abell 89B (east). Diamonds indicate galaxies detected as X-ray AGN, while
squares indicated X-ray detected galaxies that are not considered X-ray AGN.
\label{fig:x_img_85}}
\end{figure*}

\begin{figure*}
\plotone{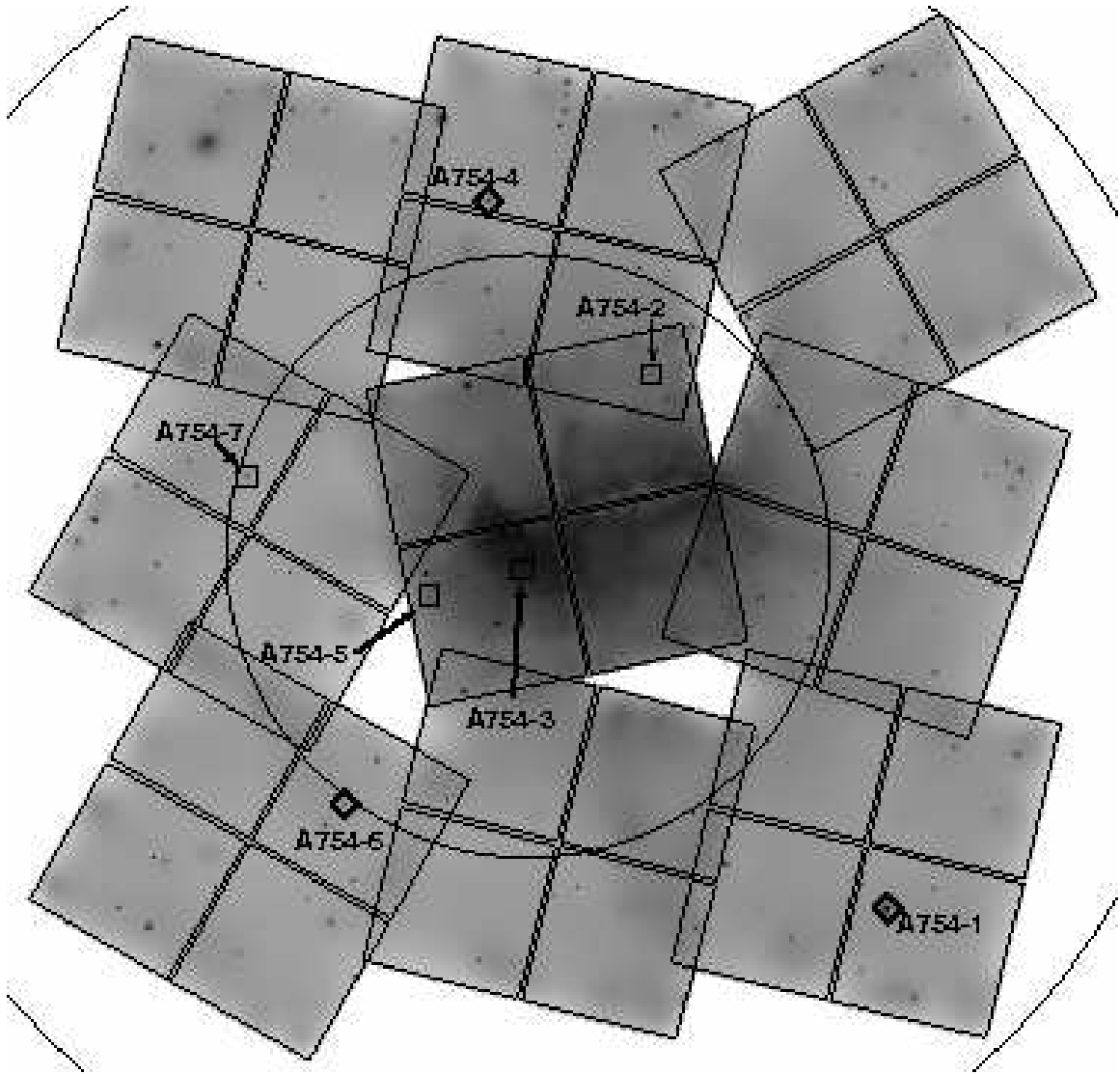}
\caption{
Adaptively smoothed {\it Chandra} mosaic of Abell 754. Overlays follow the same
conventions as Figure~\ref{fig:x_img_85}.
\label{fig:x_img_754}}
\end{figure*}

\begin{deluxetable}{lrrrc}
\tabletypesize{\footnotesize}
\tablewidth{0pt}
\tablecaption{{\it Chandra} ACIS-I Observation Logs\label{tab:xobs}}
\tablehead{
\colhead{Field} &
\colhead{OBSID} &
\colhead{Date} &
\colhead{T} &
\colhead{$L_{X,{\rm Lim}}$} \\
&
&
&
(ks) &
($10^{40} {\rm \, erg \, s}^{-1}$)\\
\colhead{(1)} &
\colhead{(2)} &
\colhead{(3)} &
\colhead{(4)} &
\colhead{(5)}
}
\startdata
Abell 85-C   & \dataset[ADS/Sa.CXO#obs/00904]{0904} & 2000-08-19 & 38.4 & $1.2$ \\
Abell 85-SE  & \dataset[ADS/Sa.CXO#obs/04881]{4881} & 2004-09-03 &  9.8 & $4.6$ \\
Abell 85-S   & \dataset[ADS/Sa.CXO#obs/04882]{4882} & 2004-09-03 &  9.6 & $4.7$ \\
Abell 85-SW  & \dataset[ADS/Sa.CXO#obs/04883]{4883} & 2004-09-03 &  9.6 & $4.7$ \\
Abell 85-E   & \dataset[ADS/Sa.CXO#obs/04884]{4884} & 2004-09-03 &  9.6 & $4.7$ \\
Abell 85-W   & \dataset[ADS/Sa.CXO#obs/04885]{4885} & 2004-09-03 &  9.6 & $4.7$ \\
Abell 85-NE  & \dataset[ADS/Sa.CXO#obs/04886]{4886} & 2004-09-03 &  9.6 & $4.7$ \\
Abell 85-N   & \dataset[ADS/Sa.CXO#obs/04887]{4887} & 2004-09-04 & 10.1 & $4.5$ \\
Abell 85-NW  & \dataset[ADS/Sa.CXO#obs/04888]{4888} & 2004-09-04 &  9.6 & $4.7$ \\
Abell 754-C  & \dataset[ADS/Sa.CXO#obs/00507]{0507} & 1999-10-30 & 29.8 & $1.3$ \\
Abell 754-SE & \dataset[ADS/Sa.CXO#obs/06793]{6793} & 2006-01-18 &  9.9 & $4.6$ \\
Abell 754-S  & \dataset[ADS/Sa.CXO#obs/06794]{6794} & 2006-01-27 &  9.9 & $4.6$ \\
Abell 754-SW & \dataset[ADS/Sa.CXO#obs/06795]{6795} & 2006-01-28 &  9.6 & $4.7$ \\
Abell 754-E  & \dataset[ADS/Sa.CXO#obs/06796]{6796} & 2006-01-18 &  9.6 & $4.7$ \\
Abell 754-W  & \dataset[ADS/Sa.CXO#obs/06797]{6797} & 2006-01-25 & 10.0 & $4.5$ \\
Abell 754-NE & \dataset[ADS/Sa.CXO#obs/06798]{6798} & 2006-01-28 & 10.0 & $4.6$ \\
Abell 754-N  & \dataset[ADS/Sa.CXO#obs/06799]{6799} & 2006-01-28 &  9.6 & $4.7$ \\
Abell 754-NW & \dataset[ADS/Sa.CXO#obs/06800]{6800} & 2006-02-15 & 10.2 & $4.5$   
\enddata
\tablecomments{{\it Chandra} ACIS-I Observation Log. Columns are:
(1) Field targeted;
(2) Observation ID of {\it Chandra} data;
(3) Observation date;
(4) Usable exposure;
(5) Estimate of the $0.3$--$8.0 {\rm \, keV}$ luminosity limit of the observation for a $z=0.055$ galaxy.}
\end{deluxetable}

\subsection{Data Reduction}

For both Abell 85 (Figure~\ref{fig:x_img_85}) and Abell 754
(Figure~\ref{fig:x_img_754}), our wide-field {\it Chandra} ACIS-I observations
consist of a $\sim 40 {\rm \, ks}$ central archival field flanked by eight new,
$\sim 10 {\rm \, ks}$ fields. We list these observations in
Table~\ref{tab:xobs}.

We reduced all data as uniformly as possible using {\sc ciao 3.4}%
\footnote{See \url{http://asc.harvard.edu/ciao/}.}
with {\sc caldb 3.3.0.1} and NASA's {\sc ftools 6.0}%
\footnote{See
\url{http://heasarc.gsfc.nasa.gov/docs/software/lheasoft/}%
\label{ftn:heasoft}.}.
Since these observations represent a combination of archival and new
observations spanning over 6 years, there were minor differences in their
reduction. For Observations 0577, 0944, and 4881-4888, the frame times were $3.2
{\rm \, s}$, while for Observations 6793-6800, the frame times were $3.1 {\rm \,
s}$. Both Observation 0577 and 0944 were telemetered and cleaned in Faint mode. 
The new observations were telemetered and cleaned in Very-Faint mode, which
leads to a reduced background. Observation 0577 was operated at $-110 ^{\circ}
\,{\rm C}$, while the remaining observations were operated at $-120 ^{\circ}
\,{\rm C}$. Thus for Observation 0577, no corrections were made for time
dependence of the gain or the charge-transfer inefficiency and photon energies
were determined using the gain file acisD1999-09-16gainN0005.fits. The other
observations were all corrected for the time dependence of the gain and the
charge-transfer inefficiency with their photon energies determined using the
gain file acisD2000-01-29gain$\_$ctiN0006.fits. For Observation 0577 and 0944, we
recreated bad pixel files using the newest tools to detect hot pixels and cosmic
ray afterglows. For all observations, we only consider events with ASCA grades
of 0, 2, 3, 4, and 6 detected by ACIS-I. Known aspect offsets were applied for
each observation. All observations were corrected for quantum efficiency
degradation and had exposure maps determined at $1.5 {\rm \, keV}$. We excluded
bad pixels, bad columns, and columns adjacent to bad columns or chip node
boundaries.

Since we use local backgrounds and small extraction regions to analyze point
sources, this analysis is not very sensitive to the periods of high background
(``background flares'') that {\it Chandra} may encounter. To avoid periods with
extreme flaring, we excluded times where the blank-sky rate was more than three
times the expected blank-sky rate derived from calibrated blank-sky backgrounds. 
We only removed $\sim 14 {\rm \, ks}$ from Observation 0507. Final
flare-filtered live exposure times for the five observations are listed in
Table~\ref{tab:xobs}.

\begin{figure}[t]
\plotone{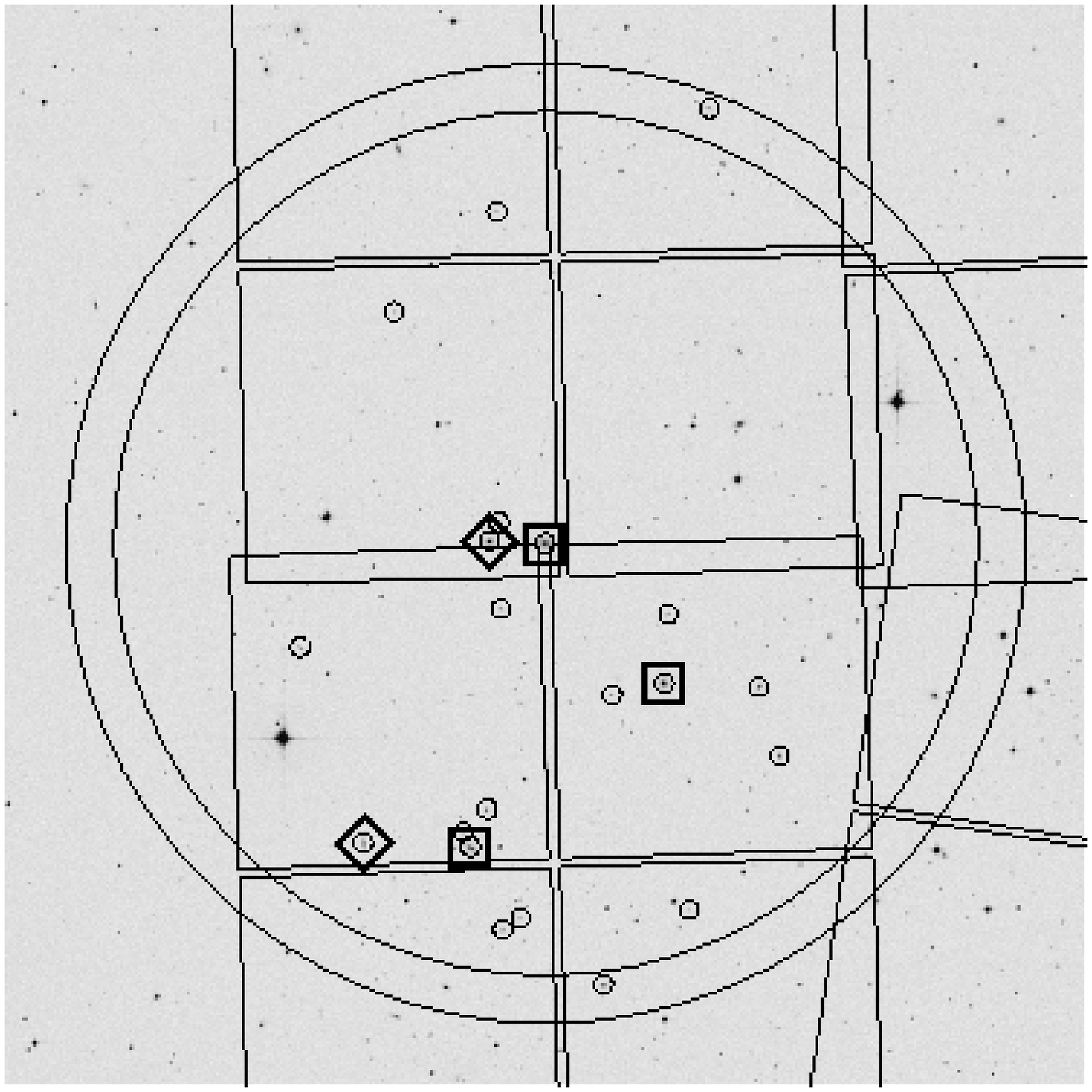}
\caption{
Second Palomar Observatory Sky Survey (Red) image centered on the BCG of
Abell 89B. Overlays follow the same conventions as Figure~\ref{fig:x_img_85},
with small circles indicating $M_R < -20$ cluster members in the {\it Chandra} FOV.
\label{fig:o_img_85}}
\end{figure}

In Figures~\ref{fig:x_img_85} and \ref{fig:x_img_754}, we display the adaptively
smoothed, exposure-corrected {\it Chandra} X-ray image of both fields using a
minimum signal-to-noise ratio (S/N) per smoothing beam of 3. The FOVs of the
individual observations are overlaid. Both clusters have ICM in the central
archival field; however, only a little diffuse gas extends into the flanking
fields. There are point sources seen in these images; however, most are
unassociated with the clusters. On these figures we also display the radii
corresponding to $1 {\rm \, Mpc}$ and $r_{200}$. Abell 85 has nearly complete
coverage to $1 {\rm \, Mpc}$ and partial coverage out to just beyond $r_{200}$. 
Although the coverage for Abell 754 is similar, there are more holes due to the
unmatched roll angles of the observations. The Abell 85 fields also provide
coverage for two other nearby large-scale structures, Abell 89B and Abell 89C
\citep{DFG+1998}. We include Abell 89B in this analysis, the less distant and
richer of the two structures. Our {\it Chandra} data covers most of Abell 89B to
its $r_{200}$ radius (Figure~\ref{fig:o_img_85}).
Abell 89C is not included as
our sample of $M_R < -20$ galaxies is incomplete at its redshift ($z \sim
0.096$) and we were unable to self-consistently identify group members using the
redshift and positions of candidate members (See \S~\ref{sec:sample}).

\subsection{Source Detection and Analysis}
\label{sec:obs_analysis}

For each observation, we applied the wavelet detection algorithm ({\sc ciao
wavdetect} program) with scales ranging from 1 to 64 pixels in steps of
$\sqrt{2}$ factors, requiring a source detection threshold of $10^{-6}$ to
identify discrete X-ray sources that are potential X-ray AGN in these clusters. 
Source detection was not performed in regions with an exposure of less than 10\%
of the total for the observation. The numbers of total detected X-ray sources are
350 and 365 in Abell 85 and Abell 754, respectively, with only a few sources
multiply detected where the FOVs overlap. Our source detection threshold
corresponds to $\la 4$ falsely detected X-ray sources (due to a statistical
fluctuation) for each observation.

There are two potential ways an X-ray source could be incorrectly associated
with an optical source: First, an associated X-ray detection could be a false
detection. Second, the positions from an X-ray detected source and an optical
counterpart could randomly overlap.
The magnitude of both effects depends on the
number of optical sources and the matching radius used to associate X-ray and
optical sources. There are 172, 21, and 270 optical members of Abell 85,
Abell 89B, and Abell 754, respectively, from \citetalias{CZ2003} in the {\it
Chandra} FOVs, and 50, 4, and 10 additional members from other sources. We first
considered a very generous $5\arcsec$ matching radius for identifying potential
X-ray emitting galaxies. This radius is large due to a $\sim3\arcsec$
uncertainty in the position of optical sources from fiber positioning
\citepalias{CZ2003} and potential poor localization of the X-ray position due to
low-count X-ray data. At this radius, we expect $\la 0.08$ and $\la 0.09$ false
associations in the Abell 85 and Abell 754 FOVs, respectively, due to
statistical fluctuations above our source detection threshold. By replacing the
source detection threshold with the average number of real X-ray sources per
pixel, we can calculate the number of false associations due to random overlap. 
We estimate $\la 0.7$ and $\la 0.9$ false associations in the Abell 85 and Abell
754 FOVs, respectively, from randomly overlapping sources. Since X-ray AGN must
be at the galaxy centers of cluster members, we apply a stricter requirement
($<2\arcsec$ offset from the 2MASS galaxy position) in \S~\ref{sec:galaxy} to
classify a source as an X-ray AGN. Thus, we estimate the expected number of
optical galaxies falsely identified as X-ray AGN is $\la 0.2$ per cluster FOV.
In addition, this expected number drops by a factor of two if we only consider
galaxies with $M_R < -20$.

\begin{deluxetable*}{llllrrr}
\tabletypesize{\footnotesize}
\tablewidth{0pt}
\tablecaption{X-ray Properties of Abell 85, Abell 89B, \& Abell 754 Galaxies
\label{tab:xgals}}
\tablehead{
\colhead{ID} &
\colhead{CXOU XID} &
\colhead{CZ2003 ID} &
\colhead{2MASX ID} &
\colhead{Offset} &
\colhead{Net Counts} &
\colhead{$L_{X,B}$} \\
\colhead{(1)} &
\colhead{(2)} &
\colhead{(3)} &
\colhead{(4)} &
\colhead{(5)} &
\colhead{(6)} &
\colhead{(7)}
}
\startdata
A85-1 & J004130.2$-$091546 & 85A\_993[6]    & J00413032$-$0915459 & $1\farcs0$ ($0\farcs3$) & $ 160.3 ^{+   15.8}_{-   14.8}$ & $  4.2^{+\phn\phn0.4}_{-\phn\phn0.4}$ \\ 
A85-2 & J004142.9$-$092621 & 85A\_993[13]   & J00414302$-$0926219 & $0\farcs8$ ($0\farcs4$) & $  21.0 ^{+\phn7.1}_{-\phn6.1}$ & $  0.5^{+\phn\phn0.2}_{-\phn\phn0.4}$ \\ 
A85-3 & J004146.7$-$092313 & 85A\_993[12]   & J00414681$-$0923129 & $0\farcs8$ ($0\farcs3$) & $   7.6 ^{+\phn4.2}_{-\phn3.0}$ & $  0.2^{+\phn\phn0.1}_{-\phn\phn0.1}$ \\ 
A85-4 & J004244.7$-$093312 & 85A\_993[86]   & J00424470$-$0933162 & $3\farcs4$ ($1\farcs1$) & $   5.8 ^{+\phn3.6}_{-\phn2.4}$ & $  0.6^{+\phn\phn0.4}_{-\phn\phn0.3}$ \\ 
A85-5 & J004311.5$-$093816 & 85A\_993[47]   & J00431162$-$0938163 & $0\farcs5$ ($0\farcs4$) & $  30.0 ^{+\phn6.5}_{-\phn5.5}$ & $  5.2^{+\phn\phn1.1}_{-\phn\phn0.9}$ \\ 
A89B-1& J004242.0$-$091731 & 85A\_993[80]   & J00424193$-$0917312 & $1\farcs7$ ($1\farcs2$) & $   7.0 ^{+\phn4.0}_{-\phn2.8}$ & $  1.5^{+\phn\phn0.9}_{-\phn\phn0.6}$ \\ 
A89B-2& J004254.8$-$091349 & 85A\_993[81]   & J00425466$-$0913493 & $2\farcs3$ ($1\farcs4$) & $  13.0 ^{+\phn4.9}_{-\phn3.8}$ & $  3.2^{+\phn\phn1.2}_{-\phn\phn0.9}$ \\ 
A89B-3& J004300.6$-$091346 & 85A\_993[57]   & J00430067$-$0913463 & $0\farcs9$ ($0\farcs6$) & $  64.0 ^{+\phn9.1}_{-\phn8.1}$ & $ 14.3^{+\phn\phn2.0}_{-\phn\phn1.8}$ \\ 
A89B-4& J004302.7$-$092151 & 85A\_993[59]   & J00430270$-$0921513 & $0\farcs5$ ($1\farcs3$) & $   3.0 ^{+\phn2.9}_{-\phn1.7}$ & $  0.6^{+\phn\phn0.6}_{-\phn\phn0.3}$ \\ 
A89B-5& J004314.0$-$092144 & 85A\_993[60]   & J00431418$-$0921453 & $1\farcs6$ ($0\farcs6$) & $   8.6 ^{+\phn4.1}_{-\phn3.0}$ & $  1.8^{+\phn\phn0.9}_{-\phn\phn0.6}$ \\ 
A754-1& J090802.1$-$095937 & 754A\_494[25]  & J09080217$-$0959378 & $0\farcs5$ ($0\farcs1$) & $1697.0 ^{+   42.3}_{-   41.3}$ & $389.7^{+      233.8}_{-      155.8}$\tablenotemark{a} \\ 
A754-2& J090852.2$-$093149 & 754A\_494[100] & J09085229$-$0931507 & $1\farcs9$ ($0\farcs9$) & $  38.8 ^{+\phn8.5}_{-\phn7.5}$ & $  1.2^{+\phn\phn0.3}_{-\phn\phn0.2}$ \\ 
A754-3& J090919.2$-$094159 & 754A\_494[9]   & J09091923$-$0941591 & $0\farcs2$ ($0\farcs4$) & $  13.5 ^{+\phn5.8}_{-\phn4.8}$ & $  0.4^{+\phn\phn0.2}_{-\phn\phn0.1}$ \\ 
A754-4& J090926.3$-$092247 & 754A\_494[93]  & J09092633$-$0922471 & $0\farcs6$ ($0\farcs3$) & $  40.4 ^{+\phn7.5}_{-\phn6.4}$ & $  4.0^{+\phn\phn0.7}_{-\phn\phn0.6}$ \\ 
A754-5& J090939.0$-$094321 & 754A\_494[106] & J09093913$-$0943233 & $3\farcs0$ ($1\farcs1$) & $  21.2 ^{+\phn7.6}_{-\phn6.6}$ & $  0.6^{+\phn\phn0.2}_{-\phn\phn0.2}$ \\ 
A754-6& J090956.8$-$095409 & 754A\_393[55]  & J09095685$-$0954093 & $0\farcs8$ ($0\farcs4$) & $  32.4 ^{+\phn6.8}_{-\phn5.7}$ & $  3.5^{+\phn\phn0.7}_{-\phn\phn0.6}$ \\ 
A754-7& J091017.3$-$093707 & 754A\_494[76]  & J09101737$-$0937068 & $1\farcs2$ ($0\farcs4$) & $  14.6 ^{+\phn5.0}_{-\phn3.9}$ & $  1.4^{+\phn\phn0.5}_{-\phn\phn0.4}$ \\ 
\enddata
\tablecomments{X-ray Measurements. Columns are:
(1) ID used in this paper;
(2) X-ray object ID;
(3) ID from \citetalias{CZ2003};
(4) 2MASS Extended Source Catalog ID of counterpart;
(5) Offset between X-ray and near-IR position with the an estimate of the $1\sigma$ statistical uncertainty of
the X-ray position in the parentheses;
(6) Net X-ray counts detected in observed frame $0.3$--$8.0 {\rm \, keV}$ band
with exact Gehrel's errors \citet{G1986};
(7) X-ray luminosity in rest frame $0.3$--$8.0 {\rm \, keV}$ band in units of
$10^{41} {\rm \, erg \, s}^{-1}$. The X-ray luminosity was calculated assuming a
$\Gamma=1.7$ power-law with corrections for Galactic absorption and the enclosed
fraction of the PSF used to extract the counts.}
\tablenotetext{a}{X-ray analysis affected by pileup. The luminosity
correction factor of $\sim2.4$ is uncertain to $\sim60\%$.}
\end{deluxetable*}

We used the coordinate list generated by {\sc wavdetect} and the positions of
optical galaxies from \citetalias{CZ2003} to identify X-ray detections within
$5\arcsec$ of optical counterparts. Due to the sensitivity of the flanking field
observations, we only considered detections and optical galaxies in regions
where the local exposure was at least half of the maximum exposure; this
eliminates the edges of the ACIS-I chips and the gaps between them. To determine
cluster membership, we adopted the velocity range in \citet{CZ2003} for Abell 85
and Abell 754. For Abell 89B, we determined its cluster properties ourselves
(see \S~\ref{sec:sample}). We found no additional matches when we added
additional cluster members from the NASA/IPAC Extragalactic Database (NED). In
Table~\ref{tab:xgals}, we list the 17 detections that correspond to a galaxy in
Abell 85, Abell 89B, or Abell 754. These galaxies are also indicated in
Figures~\ref{fig:x_img_85} and \ref{fig:x_img_754}. We label the sources in RA
order by cluster and list their X-ray position and optical counterpart from
\citetalias{CZ2003}. For each optical counterpart, we adopted the 2MASS position
in the Extended Source Catalog \citep{SCS+2006}, and recalculated the offset
between the X-ray detection and the galaxy center. Using
\citet{KKW+2007}, we have estimated the X-ray positional uncertainty (1$\sigma$)
due to {\sc wavdetect}. Our first criteria for an X-ray AGN is that the offset
between the X-ray detection and the galaxy is less than $2\arcsec$, consistent
with that used in \citetalias{MKK+2006}. Since all three detections that fail
this criterion have $\sim 1\arcsec$ positional uncertainty, they are still likely
associated with the identified galaxies. 
We have excluded a detection consistent
with the brightest cluster galaxy (BCG) of Abell 85, as this detection also
corresponds to the peak in the X-ray flux from ICM.
We also note that a
detection corresponding to an Abell 85 member that is likely an X-ray AGN with
$L_{X,B} \sim 1.2 \times 10^{41} {\rm \, erg \, s}^{-1}$ (2MASX J00415019-0925469) was
excluded since it fell in a chip gap of Observation 0904 and the photometry is
therefore highly uncertain.

\begin{deluxetable*}{lrrrrrrc}
\tabletypesize{\footnotesize}
\tablewidth{0pt}
\tablecaption{Cluster Properties
\label{tab:cluster}}
\tablehead{
\colhead{Cluster} &
\colhead{$\alpha_c$} &
\colhead{$\delta_c$} &
\colhead{$z$} & 
\colhead{$z_1$,$z_2$}&
\colhead{$\sigma$}&
\colhead{$r_{200}$}&
\colhead{Reference} \\
\colhead{(1)}&
\colhead{(2)}&
\colhead{(3)}&
\colhead{(4)}&
\colhead{(5)}&
\colhead{(6)}&
\colhead{(7)}&
\colhead{(8)}
}
\startdata
Abell 85   & 00:41:50.4 & $-09$:18:11 & 0.0554    & 0.0448,0.0658 & 993 (\phn85)  & 2.4 & 1 \\
Abell 89B  & 00:42:54.6 & $-09$:13:50 & 0.077\phn & 0.0692,0.0850 & 474 (155)     & 1.1 & 2 \\
Abell 3125 & 03:25:17.9 & $-53$:29:37 & 0.0616    & 0.0530,0.0700 & 475 (\phn94)  & 1.1 & 3 \\
Abell 3128 & 03:30:43.8 & $-52$:31:30 & 0.0595    & 0.0435,0.0755 & 906 (\phn74)  & 2.1 & 3 \\
Abell 754  & 09:09:18.0 & $-09$:41:17 & 0.0546    & 0.0446,0.0632 & 953 (\phn64)  & 2.3 & 1 \\
Abell 644  & 08:17:25.6 & $-07$:30:45 & 0.0701    & 0.0531,0.0871 & 952 (382)     & 2.2 & 3 \\
\enddata
\tablecomments{
Sample of $z \lesssim 0.08$ clusters with X-ray identified AGN. Columns are: (1) Cluster
name; (2 and 3) RA and DEC of the cluster center for epoch J2000; (4) Redshift;
(5) Redshift range of cluster members; (6) Velocity dispersion and uncertainty (90\%);
(7) $r_{200}$ in Mpc;
(8) Reference for velocity information.
}
\tablerefs
{
(1) \citetalias{CZ2003};
(2) this paper;
(3) \citetalias{MMK2007};
}
\end{deluxetable*}

For all detections in Table~\ref{tab:xgals}, we used ACIS Extract 3.131 to
create source extraction regions enclosing 90\% of the flux in the X-ray PSF and
to determine a masking radius that encircled 97\% of the flux. For most of the
sources, whose photons had median energies of $\sim 0.6$--$2.6 {\rm \, keV}$, we
determined the regions assuming the PSF at $1.497 {\rm \, keV}$. Since the
events for A754-6 had a median energy of $\sim 4.7 {\rm \, keV}$, we used the
PSF determined at $4.51 {\rm \, keV}$. For each source, we created background
regions just beyond the masking radius with an area five times that of the
source extraction region. Column (6) of Table~\ref{tab:xgals} indicates the net
counts for each source in the $0.3$--$8.0 {\rm \, keV}$ band, with proper
Poisson errors \citep{G1986}. To estimate the rest-frame $0.3$--$8.0 {\rm \,
keV}$ X-ray luminosity, column (7), we folded a power-law spectrum with
$\Gamma=1.7$ absorbed by the Galactic column ($3.3\times10^{20} {\rm cm}^2$ for
Abell 85 and Abell 89B; $4.4\times10^{20} {\rm cm}^2$ for Abell 754) through the
spectral response at the location of each source. We set the model normalization
using {\sc xspec}\footnotemark[\ref{ftn:heasoft}] to match the observed net
counts, corrected for the mean redshift of the cluster and the enclosed fraction
of the flux in the source extraction region.

We note that A754-1 is bright enough that it suffers from events lost to pileup. 
At $\sim 0.55$ counts per frame, pileup can be relatively minor and require only
a small correction or pileup can be more severe and require a larger correction. 
Since there is no readout-streak and the source is shaped like the PSF, the
branch with less pileup is more likely correct. Therefore, the luminosity in
Table~\ref{tab:xgals} has been corrected assuming that our spectral model has
been affected by pileup with a typical grade-migration parameter, $\alpha=0.5$
\citep{D2001}. We estimate that the correction factor of $\sim2.4$ is accurate
to a factor of $\sim 60\%$. If the source is more extremely effected by pileup,
this will only increase its X-ray luminosity.

By combining the luminosities and counts from Table~\ref{tab:xgals} with our
exposure maps, we estimated the limiting X-ray luminosity for each observation. 
This is listed in Table~\ref{tab:xobs} assuming a redshift of $z=0.055$ near
that of Abell 85 and Abell 754. For the more distant Abell 89B, the limiting
luminosity is a factor of two higher. For consistency with
\citetalias{MKK+2006}, we have calculated this number corresponding to five
counts on-axis. 
However, we caution that this limit is optimistic over an entire
ACIS-I FOV for two reasons. First, at $1.5 {\rm \, keV}$ the spatial structure
to the quantum efficiency degradation leads to 15\% lower exposure at
approximately $10\arcmin$ off-axis in the latest observations. More importantly,
the larger off-axis PSF makes detection of weak sources more difficult. 
\citet{KF2003} show that at $5\arcmin$ and $10\arcmin$ off-axis, 70\%
completeness can be expected for 7 and 11 counts sources respectively.
We estimate that completeness limits over the entire ACIS-I FOV are about a
factor of four higher than reported in Table~1 of \citetalias{MKK+2006} and
Table~\ref{tab:xgals} of this work. 
This means that the AGN fraction above $10^{41} {\rm \, erg \, s}^{-1}$ may be
underestimated; however, we estimate that this is a smaller effect than the
current error due to the small numbers of AGN.

\section{$z \lesssim 0.08$ Cluster Sample}
\label{sec:sample}

We required a sample large enough to statistically test which galaxy and cluster
properties lead to X-ray AGN activity. This is especially important as not all
detected X-ray sources will be X-ray AGN. To supplement the sample of 17
potential X-ray AGN in Abell 85, Abell 89B, and Abell 754, we have also included
three other $z \lesssim 0.08$ clusters with
X-ray identified AGN, Abell 644, Abell 3125, and Abell 3128
\citepalias{MKK+2006,MMK2007}.
We list the cluster properties in
Table~\ref{tab:cluster}, adopting the \citetalias{MMK2007} values for the latter
three clusters.

In columns (2) and (3) , we list the cluster positions. For Abell 85 and Abell
754, we adopted the peak of the ICM as the cluster position. The BCG of Abell 85
is coincident with this peak. In Abell 754, the third brightest galaxy (in
$R$-band), A754-3, is embedded in the ICM $\sim46\arcsec$ away from this
position. This galaxy is located near one of the concentrations of Abell 754
member galaxies.

We list the mean cluster redshift, redshift range of cluster members, and
velocity dispersion, with 90\% confidence limit, in columns (4)--(6). We adopted
the values of \citetalias{CZ2003} for Abell 85 and Abell 754; however, we
increased their $1\sigma$ uncertainties in velocity dispersion by a factor of
1.6 to match our confidence limits. For Abell 89B, the
\citet{CZ2003} data suggested that its members were in the
$0.06 < z < 0.09$ range. We calculated membership via the biweight estimator for
center and scale, following \citetalias{MMK2007}, adding additional nearby
galaxies with velocity data in the NED to the \citetalias{CZ2003} sample. We
iteratively determined 29 galaxies were within 5$\sigma$ of the cluster mean
velocity and the $r_{200}$, assuming the BCG was the center of the cluster. Of
the 29 galaxies, our FOV overlapped with 25. We used the jackknife of the
biweight estimator to determine the 90\% confidence limit for the velocity
dispersion. The symmetric confidence limit \citep[eq.\ 22 of ][]{BFG1990} was
chosen for consistency with \citetalias{MKK+2006,MMK2007}.

To characterize the extent of the clusters and best compare the spatial
distributions of cluster AGN, we determined the $r_{200}$ of each cluster
following equation A1 of \citet{TEK+2003}. These are listed in column (7).

\section{Galaxy Properties of X-ray Sources}
\label{sec:galaxy}

\subsection{X-ray AGN Identification}

Near the luminosity limits of these observations, there are three potential
sources of X-ray emission: X-ray binaries, hot ISM, and a central AGN
\citep[e.g.,][]{SSI2003,SSC2004}. X-ray
binaries with low-mass companions (LMXBs) are sensitive to the total stellar
mass of a galaxy, while X-ray binaries with high-mass companions (HMXBs) are
sensitive to recent star formation
\citep{GGS2003,KF2004}.
From a sample of fourteen nearby galaxies,
\citet{KF2004} derived a (linear) relation between the total X-ray luminosity of
LMXBs within the galaxy and the $B$-band or $K_s$-band luminosity. We prefer the
latter relation as $K_s$-band is a better tracer of stellar mass, and the
relation has a smaller dispersion;
\begin{equation}
\label{eq:l_x_lmxbs}
L_{X,B} = (2.0 \pm 0.8) \times 10^{29} {\rm \, erg \, s}^{-1}
          / L_{K_{s},K20,\odot},
\end{equation}
where $L_{K_{s},K20,\odot}$ is the $K_s$-band luminosity within the $K_s = 20
{\rm \, mag \, arcsec}^2$ isophote, assuming $M_{\odot,K_s} = 3.33$.
We caution that most of these galaxies in this archival sample were originally
targeted due to their X-ray properties. These galaxies are roughly divided into
X-ray bright galaxies, galaxies with significantly higher X-ray to optical flux
ratios that are dominated by the diffuse gas, and X-ray faint galaxies, galaxies
with lower X-ray to optical flux ratios that are dominated by the X-ray
binaries. The X-ray bright galaxies should be relatively free from a bias on the
total LMXB X-ray luminosities in these systems. As studying LMXBs was often the
primary science driver for targeting the X-ray faint galaxies, these galaxies
were often selected
based on their X-ray luminosities or X-ray
to optical flux ratios. Due to such selection criteria, X-ray faint galaxies
whose X-ray luminosities are towards the lower-end of the intrinsic relation
between X-ray luminosity from LMXBs and stellar mass are less likely to be
targeted by observers. Thus, the above relation may overestimate the intrinsic
relation.
To estimate the X-ray luminosity from HMXBs, the star-formation rate
(SFR) is needed. Assuming a $\Gamma=1.7$ X-ray spectrum, we can convert the
relation found in
\citet{GGS2003} to a $L_{X,B}$, such that
\begin{equation}
L_{X,B} = 1.0 \times 10^{40}
          \frac{{\rm SFR}}{{\rm M}_\odot {\rm \, yr}^{-1}}
          {\rm \, erg \, s}^{-1}. 
\end{equation}

Since the ISM is thought to have a stellar origin, a rough
correspondence with stellar mass is expected; however, at a given stellar mass
there is a wide-range of ISM luminosities and the relation to stellar mass is
known to be non-linear. We adopt the \citet{SJF+2007} relation
\begin{equation}
\label{eq:l_x_gas}
\log L_{X,S} = 39.40 + (1.63\pm0.13) \log \left(\frac{L_{K_{s},{\rm ttl},\odot}}{10^{11}}\right),
\end{equation}
where the soft ($0.5$--$2.0 {\rm \, keV}$) band X-ray luminosity,
$L_{X,S}$ is calculated assuming an ISM spectral model and
$L_{K_{s},{\rm ttl},\odot}$ is the total $K_s$-band luminosity. 
This relation is derived including the effects of upper limits for
non-detections of the ISM. For the galaxies in our cluster sample, we have
estimated that $L_{K_{s},{\rm ttl}} \sim 1.23 L_{K_{s},K20}$ and that the
$L_{X,B}$ for a $\Gamma =1.7$ power-law is $\sim 1.9$ times the $L_{X,S}$ for
$kT = 0.7 {\rm \, keV}$ gas with 0.8 solar abundance when requiring that the
observed $0.3$--$8.0 {\rm \, keV}$ count-rates match.
Note that applying just a
luminosity cut of $L_{X,B} = 10^{41} {\rm \, erg \, s}^{-1}$ to identify AGN can
be contaminated by galaxies without AGN if either $L_{K_s,\odot} \gtrsim 2.5
\times 10^{11}$ or ${\rm SFR}
\gtrsim 10 {\rm \, M}_\odot {\rm \, yr}^{-1}$.

\begin{deluxetable}{lrrrr}
\tabletypesize{\scriptsize}
\tablewidth{0pt}
\tablecaption{Optical/Near-IR Magnitudes of X-ray Galaxies
\label{tab:ogals}}
\tablehead{
\colhead{ID} &
\colhead{$m_{R}$} &
\colhead{$M_{R}$} &
\colhead{$M_{K_s,K20}$} &
\colhead{Ref.} \\
\colhead{(1)} &
\colhead{(2)} &
\colhead{(3)} &
\colhead{(4)} &
\colhead{(5)}
}
\startdata
A85-1\tablenotemark{a}   & 14.32 & $-22.81$ & $-25.14$ & 1 \\
A85-2                    & 14.43 & $-22.70$ & $-25.43$ & 1 \\
A85-3                    & 15.86 & $-21.27$ & $-24.05$ & 1 \\
A85-4                    & 14.81 & $-22.10$ & $-24.98$ & 1 \\
A85-5 \tablenotemark{a}  & 15.03 & $-22.22$ & $-24.75$ & 1 \\
A89B-1                   & 14.18 & $-23.08$ & $-25.81$ & 1 \\
A89B-2                   & 15.03 & $-23.71$ & $-26.37$ & 1 \\
A89B-3\tablenotemark{a}  & 15.17 & $-22.85$ & $-25.51$ & 1 \\
A89B-4                   & 14.92 & $-22.72$ & $-25.81$ & 1 \\
A89B-5\tablenotemark{a}  & 14.09 & $-22.03$ & $-25.19$ & 1 \\
A754-1\tablenotemark{a}  & 14.40 & $-22.79$ & $-25.69$ & 1 \\
A754-2                   & 14.60 & $-22.54$ & $-25.30$ & 1 \\
A754-3                   & 14.05 & $-23.13$ & $-25.63$ & 1 \\
A754-4\tablenotemark{a}  & 14.24 & $-22.90$ & $-25.79$ & 1 \\
A754-5                   & 14.28 & $-22.87$ & $-25.60$ & 1 \\
A754-6\tablenotemark{a}  & 15.84 & $-21.33$ & $-24.10$ & 1 \\
A754-7                   & 13.93 & $-23.19$ & $-26.00$ & 1 \\
A3125-1\tablenotemark{a} & 15.77 & $-21.56$ & $-24.66$ & 2 \\
A3125-2                  & 14.93 & $-22.39$ & $-25.62$ & 2 \\
A3125-3                  & 15.24 & $-22.08$ & $-25.43$ & 2 \\
A3125-4                  & 15.16 & $-22.17$ & $-25.42$ & 2 \\
A3125-5\tablenotemark{a} & 15.97 & $-21.36$ & $-24.54$ & 2 \\
A3125-6\tablenotemark{a} & 14.97 & $-22.36$ & $-24.18$ & 3 \\
A3128-1                  & 15.24 & $-22.01$ & \nodata  & 2 \\
A3128-2\tablenotemark{a} & 17.17 & $-20.08$ & \nodata  & 2 \\
A3128-3                  & 16.21 & $-21.04$ & $-24.16$ & 4 \\
A3128-4\tablenotemark{a} & 14.81 & $-22.43$ & $-26.09$ & 2 \\
A3128-5                  & 15.65 & $-21.60$ & $-24.62$ & 5 \\
A3128-6\tablenotemark{a} & 16.82 & $-20.43$ & $-23.22$ & 2 \\
A3128-7                  & 15.01 & $-22.23$ & $-25.77$ & 2 \\
A3128-8                  & 15.28 & $-21.97$ & $-26.02$ & 2 \\
A3128-9\tablenotemark{a} & 16.41 & $-20.83$ & $-23.76$ & 2 \\
A3128-10                 & 14.60 & $-22.65$ & $-26.33$ & 2 \\
A644-1\tablenotemark{a}  & 16.63 & $-21.94$ & $-24.80$ & 2 \\
A644-2\tablenotemark{a}  & 15.90 & $-21.23$ & $-24.22$ & 2 \\
\enddata
\tablecomments{Optical/Near-IR Measurements of X-ray Identified Galaxies in Six $z<0.08$ Clusters.
Columns are:
(1) ID from this paper or \citetalias{MKK+2006};
(2) Observed $R$-band magnitude;
(3) Extinction corrected rest-frame absolute $R$-band magnitude;
(4) Extinction corrected rest-frame absolute $K_s$-band magnitude within the $K_s = 20 {\rm
    \, mag \, arcsec}^2$ isophote;
(5) Reference for $R$-band magnitude
}
\tablenotetext{a}{Galaxy selected as X-ray AGN.}
\tablerefs
{
(1) \citetalias{CZ2003};
(2) \citetalias{MKK+2006};
(3) \citealt{LV1989};
(4) \citealt{CR1997};
(5) \citealt{KMH+1998}
}
\end{deluxetable}

In Table~\ref{tab:ogals}, we list the optical/near-IR magnitudes for galaxies in
our sample of clusters. In column (2), we list the observed $R$-band magnitude. 
We list the references for these magnitudes in column (5). The absolute $R$-band
magnitude, including extinction corrections \citep[$A_R=2.64 \
\nobreak{E(B-V)}$;][]{SFD1998}, are listed in column (3), assuming a distance
corresponding to the mean redshift of each cluster.
As in \citet{MKK+2006}, we applied corrections for
bandpass shifting and stellar evolution based on a simple stellar population
model with solar metallicity and formation redshift of $z=3$ \citep{BC2003}. 
At these redshifts the corrections to the $R$-band magnitudes are small
(0.06--0.08).
All X-ray sources are in
galaxies with $M_R < -20$.
For comparison, we note that the knee of local galaxy luminosity functions occurs at
$M^*_R = -21.15$ \citepalias{CZ2003}. In column (4), we list the absolute $K_s$-band
magnitude with extinction corrections \citep[$A_{K_s}=0.28 \
\nobreak{E(B-V)}$;][]{MSW+2003}, where we have used the 2MASS magnitude within
the $K_s = 20 {\rm \, mag \, arcsec}^2$ isophote
\citep{SCS+2006}.
The correction for bandpass shifting and stellar evolution to the $K$-band magnitudes
are larger (0.24--0.32) than those applied to the $R$-band magnitudes.
While we do not have robustly measured SFR for these galaxies, we place rough
limits on the SFR in \S~\ref{sec:SFR}.

\begin{figure}
\plotone{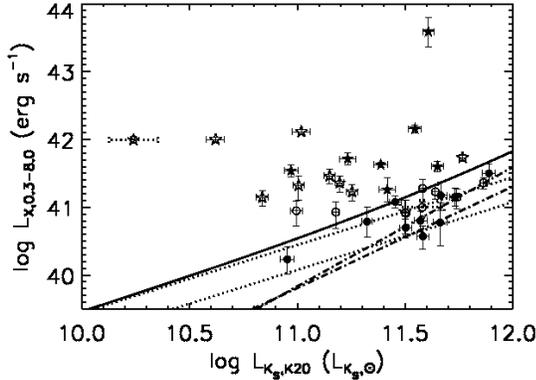}
\caption{
Broad band X-ray luminosity, $L_{X,B}$, versus the near-IR luminosity enclosed
in the $K_s = 20 {\rm \, mag \, arcsec}^2$ isophote, $L_{K_{s},K20}$, for X-ray
detected galaxies in the cluster sample from Table~\ref{tab:cluster}. The
$1\sigma$ range of X-ray emission expected from LMXBs \citep[{\itshape dotted}
line,][]{KF2004} and diffuse gas \citep[{\itshape dash-dotted} line,][]{SJF+2007} are
displayed. Galaxies that have $L_{X,B}$ brighter than $10^{41} {\rm \, erg \,
s}^{-1}$ and more than $1\sigma$ away from the sum of the upper limits for LMXBs
and diffuse gas ({\itshape solid} line are considered X-ray AGN and are marked
by stars. Filled and open symbols indicate galaxies from this paper and
\citet{MKK+2006}, respectively. Two of the galaxies from \citetalias{MKK+2006}
had no 2MASS Extended Source Catalog counterpart and have estimated
$L_{K_{S,K20}}$ and larger errors ({\it thick dotted} bars). The most luminous X-ray
source, A754-1, has been corrected for pileup, which is uncertain to $\sim60\%$.
\label{fig:lx_lk}}
\end{figure}

In Figure~\ref{fig:lx_lk}, we plot the X-ray luminosity versus the $K_s$-band
luminosity for galaxies in our cluster sample. The errors for the X-ray
luminosity are calculated from the errors in the count-rates alone, except for
A754-1 whose errors arise from uncertainty in the pileup correction. To
estimate the near-IR luminosity for the two galaxies that were not in the 2MASS
Extended Source Catalog, we used the relation between the standard aperture
$K_s$ magnitude in the Point Source Catalog and the $K_s$-band isophotal
magnitude for the other galaxies. These two galaxies are indicated with their
larger {\it dashed} error bars. Galaxies with X-ray luminosities newly measured
by this paper are indicated with filled symbols in Figure~\ref{fig:lx_lk}. We
overlay the $1\sigma$ ranges of the \citet{KF2004} and \citet{SJF+2007}
relations after correcting the latter to isophotal optical luminosities and
$L_{X,B}$ assuming a $\Gamma=1.7$ power-law. The {\it solid} line indicates the
sum of the upper limits from both relations.

We classify a galaxy as an X-ray AGN if the following conditions are met:
$L_{X,B} > 10^{41} {\rm \, erg \, s}^{-1}$,
$L_{X,B}$ more than $1\sigma$ higher than the sum of the $1\sigma$ upper limits to the
\citet{KF2004} and \citet{SJF+2007} relations, and an optical counterpart within
$2\arcsec$. These galaxies are indicated by a note in
Table~\ref{tab:ogals} and with a star in Figure~\ref{fig:lx_lk}.
One source is marginally above the sum of the $1.3\sigma$ upper limits to the
\citet{KF2004} and \citet{SJF+2007} relations, A89B-5; all other sources are
above the sum of the $2.7\sigma$ upper limits of the relations.
Since our X-ray luminosity is
derived for a point-source, and not the entire galaxy, we note that the total
galaxy X-ray luminosity will be even larger than that in Figure~\ref{fig:lx_lk}
if there is a contribution from the extended emission of the distribution of
LMXBs or ISM. Thus, the only likely contaminating sources in this sample of
X-ray AGN are galaxies with ${\rm SFR}
\gtrsim 10 {\rm \, M}_\odot {\rm \, yr}^{-1}$. We argue below that such
contamination does not seem likely for our sample. 
We also note that A3128-3 is an X-ray AGN if we do not impose an X-ray
luminosity cut, i.e., it has a close optical counterpart, is above the sum of
the LMXB and ISM relations, but has $L_{X,B} < 10^{41} {\rm \, erg \, s}^{-1}$.

Although one could consider adding additional requirements to classify a source
as an X-ray AGN based on its X-ray data, in particular estimates of its spectrum
(e.g., through hardness ratios or quantiles) and spatial decomposition into a
point source and extended galactic emission, the quality of the data for the
lower luminosity sources is insufficient. First, only 20\% of the X-ray detected
galaxies have more than 50 counts.
There would be little to no discriminating
power for the vast majority of our sample.
Second, it is unclear that a
spectral selection using hardness ratios or quantiles is appropriate.
\citetalias{MKK+2006} found that the spectroscopically identified AGNs were those
least consistent with unobscured, $\Gamma = 1.7$, power-law emission. While one
might hope to discriminate the soft emission of diffuse gas from harder
power-law emission, some AGNs have ultrasoft spectra, that corresponds to steep
power-law photon indices, $\Gamma \gtrsim 3$ \citep{PMC+1992}. This highlights
the need for deep enough observations where spectral modeling can be done to
detect the iron L-shell hump characteristic of diffuse gas
\citep[e.g.,][]{SJF+2007}.
Since most of our X-ray detected galaxies have less than 50 counts, 
spatial decomposition of the X-ray emission would also not be useful
for the vast majority of galaxies in our sample.

Based on \citet{KF2003}, we estimate that the completeness limits over the
entire ACIS-I FOV is approximately four times the X-ray luminosity in Table~1 of
\citetalias{MKK+2006} and Table~\ref{tab:xgals} of this work.
This suggests that the central observation of Abell 85 and Abell 754 are
incomplete at $L_{X,B} \lesssim 5 \times 10^{40} {\rm \, erg \, s}^{-1}$, while the
flanking field observations are incomplete for $L_{X,B} \lesssim 2\times 10^{41} {\rm
\, erg \, s}^{-1}$. Since Abell 89B is in flanking field observations of Abell
85 and is more distant, it is incomplete for $L_{X,B} \lesssim 4\times 10^{41} {\rm \,
erg \, s}^{-1}$. Abell 644 and Abell 3128 are incomplete for $L_{X,B} \lesssim 10^{41}
{\rm \, erg \, s}^{-1}$, while Abell 3125 is incomplete for $L_{X,B} \lesssim 2\times
10^{41} {\rm \, erg \, s}^{-1}$. Although there is a gap between $10^{41}
{\rm \, erg \, s}^{-1}$ and the completeness limits in some areas of the clusters, we
estimate that the completeness in this gap is above 50\%.
Since only one X-ray AGN is detected in the gap between
$L_{X,B} > 10^{41} {\rm \, erg \, s}^{-1}$ and its completeness limit, A89B-5,
we estimate that we are not likely to be missing more than one or two X-ray AGN
due to incompleteness.

\subsection{Spectroscopically Identified AGN in Abell 85, Abell 89B, and Abell 754}

\begin{deluxetable}{lrrr}
\tabletypesize{\footnotesize}
\tablewidth{0pt}
\tablecaption{Optical Spectral Properties of X-ray Galaxies
\label{tab:spgals}}
\tablehead{
\colhead{ID} &
\colhead{EW [\ion{O}{2}]} &
\colhead{EW [\ion{O}{3}]} &
\colhead{EW H$\beta$}\\
\colhead{(1)} &
\colhead{(2)} &
\colhead{(3)} &
\colhead{(4)}
}
\startdata
A85-1 &$37.88\pm2.31$&$84.88\pm3.20$&$ 4.61\pm5.04$\\
A85-2 &$-0.21\pm0.69$&$ 0.15\pm0.32$&$-0.41\pm0.18$\\
A85-3 &$ 1.53\pm1.11$&$ 0.02\pm0.51$&$-0.29\pm0.31$\\
A85-4 &$ 4.06\pm2.53$&$ 0.21\pm0.25$&$-0.08\pm0.28$\\
A85-5 &$ 1.08\pm1.04$&$-0.36\pm0.48$&$-0.34\pm0.27$\\
A89B-1&$-0.49\pm0.59$&$ 0.14\pm0.51$&$-0.03\pm0.22$\\
A89B-2&$ 3.30\pm1.90$&$-0.29\pm0.28$&$ 0.00\pm0.25$\\
A89B-3&$ 1.29\pm1.03$&$-0.15\pm0.24$&$-0.59\pm0.19$\\
A89B-4&$ 4.29\pm1.79$&$ 1.40\pm0.66$&$-0.09\pm0.26$\\
A89B-5&$21.99\pm2.45$&$25.47\pm1.65$&$-0.11\pm0.63$\\
A754-1&$ 3.41\pm2.06$&$-0.18\pm0.30$&$-0.25\pm0.26$\\
A754-2&$ 4.24\pm2.38$&$-0.23\pm0.30$&$-0.34\pm0.22$\\
A754-3&$-0.30\pm0.64$&$-0.10\pm0.16$&$-0.39\pm0.13$\\
A754-4&$ 1.51\pm1.20$&$ 0.36\pm0.46$&$-0.22\pm0.19$\\
A754-5&$-0.42\pm0.60$&$ 0.28\pm0.32$&$-0.10\pm0.15$\\
A754-6&$10.36\pm1.42$&$11.05\pm1.37$&$ 0.20\pm0.55$\\
\enddata
\tablecomments{Optical Spectral Properties of X-ray Identified Galaxies in Abell 85, Abell 89B, and Abell 754.
Columns are:
(1) ID from this paper;
(2) Equivalent width of [\ion{O}{2}] emission;
(3) Equivalent width of [\ion{O}{3}] emission;
(4) Equivalent width of H$\beta$ emission without correction for absorption.
Abell 754-7 is not included due to its spectrum having low signal-to-noise.
}
\end{deluxetable}

In addition to measuring the redshifts of galaxies in Abell 85, Abell 89B, and
Abell 754, the spectroscopy described in \citetalias{CZ2003} and
\citet{CZ2005} yielded measurements of the equivalent widths of
the [\ion{O}{2}] $\lambda 3727$ doublet, [\ion{O}{3}] $\lambda 5007$, and
H$\beta$\ $\lambda 4861$ emission lines.
The last is not corrected for any H$\beta$ absorption.
These emission lines can be indicative of ionization from an AGN and/or current,
unobscured star-formation. These values are listed in Table~\ref{tab:spgals} for
X-ray detected galaxies. We used our spectroscopic measurements to check for AGN
identifiable by their optical spectra among our X-ray detected galaxies in Abell
85, Abell 89B, and Abell 754. The spectroscopic identification of AGN in our
other clusters was previously discussed in \citetalias{MKK+2006}.

Only three of the X-ray detected galaxies, A85-1, A89B-5, and A754-6, show
emission lines detected at $>3\sigma$. All three galaxies, which are classified
as X-ray AGN, have significant detections of [\ion{O}{2}] and [\ion{O}{3}];
however, none of them have H$\beta$ emission. To conservatively correct for
potential absorption, we have added the emission-corrected H$\beta$ absorption
equivalent widths of 5 \AA\ found for post-starburst galaxies in the Sloan
Digital Sky Survey (C. Tremonti, private communication) to all measurements of
H$\beta$. Both A85-1 and A89B-5 are spectroscopically classified as AGN via the
[\ion{O}{3}]/H$\beta$ versus [\ion{O}{2}]/H$\beta$ diagnostic
\citep{RTT1997,LMC+2004};
\begin{equation}
\log \left(\frac{[\mion{O}{3}]}{{\rm H}\beta}\right) > 
  \frac{0.14}{[\mion{O}{2}]/{\rm H}\beta-1.45}+0.83.
\end{equation}
No other galaxies in the {\it Chandra} FOVs of these clusters are
spectroscopically identified as AGN with our emission line data.
Since A85-1 has been previously identified as a Seyfert galaxy
\citep{DLF2005} and A89B-5 has been classified as a QSO by SDSS
\citep{SDSS_DR5}, their identifications as spectroscopically identified AGN
appear secure.

In addition to the galaxies we identify as spectroscopic AGN, two
other X-ray detected galaxies have been previously identified as AGN based on
their optical spectral properties. A85-2 was identified as a Seyfert 2
\citep{HB1991}; however, we note that the redshift associated with this
identification (0.0453) does not match our measured redshift of the galaxy
(0.0564). A85-3 was identified as an AGN based on the limit to its
[\ion{N}{2}]/H$\alpha$ ratio \citep{HHP+2005}. Both A85-2 and A85-3 were not
identified as X-ray AGN because their low X-ray luminosities ($L_{X,B} < 10^{41}
{\rm \, erg \, s}^{-1}$) were consistent with emission from their LMXB
populations. These sources illustrate that although current X-ray observations
allow identification of low-luminosity AGN, some lower-luminosity AGN are still
being missed. Another famous example is that the X-ray emission from the core
and jet of M87 \citep[e.g.,][]{MMD+2002} would not be luminous enough to be
classified as an X-ray AGN with our criteria.

\subsection{Star Formation Rates}
\label{sec:SFR}

Since HMXBs associated with star formation may also lead to
X-ray emission, it is important to evaluate whether an X-ray detected galaxy has
a high SFR. We use optical spectroscopy to constrain the current SFR through the
[\ion{O}{2}] equivalent width. Rough limits on the SFR for X-ray galaxies can be
estimated from detections and limits based on the Infrared Astronomy Satellite
{\it IRAS} Faint Source Catalog \citep{M+1990}. For Abell 85, Abell 89B, Abell
754, and Abell 644, radio fluxes and limits from the NRAO VLA Sky Survey
\citep[NVSS,][]{CCG+1998} are also available to constrain the current SFR.

In the absence of an AGN component, the [\ion{O}{2}] equivalent width can be
used to estimate the SFR that is unobscured,
\begin{equation}
{\rm SFR}_{[{\rm O \, II}]} = 8.8 \times 10^{-12} \ L_{B,\odot} \,
                              {\rm EW}[\mion{O}{2}]
                              {\rm \, M}_\odot {\rm \, yr}^{-1},
\end{equation}
where $L_{B,\odot}$ is the $B$-band luminosity in solar luminosities
\citep{K1992b,BP1997}.
Among X-ray sources in Abell 644, Abell 3125, and Abell 3128, only A3125-5 and
A3128-2 have measurable [\ion{O}{2}] emission \citepalias{MKK+2006}; however,
the implied SFR$_{[{\rm O \, II}]}$ for both sources is small ($\lesssim 1 {\rm
\, M}_\odot {\rm \, yr}^{-1}$). For galaxies in Abell 85, Abell 89B, and Abell
754, we estimate $L_{B,\odot}$ assuming $B-R = 1$, which is appropriate for
cluster X-ray sources with [\ion{O}{2}] emission \citepalias{MKK+2006}. Among
the galaxies with $3\sigma$ detections of [\ion{O}{2}], two have SFR$_{[{\rm O
\, II}]} > 5 {\rm \, M}_\odot {\rm \, yr}^{-1}$, A85-1 ($26 {\rm \, M}_\odot
{\rm \, yr}^{-1}$) and A89B-5 ($7 {\rm \, M}_\odot {\rm \, yr}^{-1}$). A85-1 is
an Sbc galaxy \citep{PPP+2003} whose peculiar velocity ($-3.2$ times the
velocity dispersion of Abell 85) suggests it is an infalling galaxy towards the
edge of the Abell 85 despite its small projected cluster-centric distance ($0.15
{\rm \,} r_{200}$). A89B-5 is also a late type-galaxy \citep[S?;][]{PPP+2003}
at the edge of Abell 89B ($0.73 {\rm \,} r_{200}$). If the [\ion{O}{2}]
equivalent widths of A85-1 and A89B-5 were indicative of their SFR, then
approximately 60\% and 40\% of their X-ray emission could come from HMXBs. 
However, our identification of both as spectroscopic AGN suggests their
[\ion{O}{2}] likely includes a considerable AGN component. This would lead to an
overestimate of their SFR$_{[{\rm O \, II}]}$ and implied HMXB X-ray luminosity.

For galaxies without $3\sigma$ detections of [\ion{O}{2}], we conservatively
adopted three times the measurement error of [\ion{O}{2}]. We have excluded
A754-7 because its spectra had low signal-to-noise. Only one remaining galaxy
had a large implied SFR$_{[{\rm O \, II}]}$, A89B-2 ($<8.8 {\rm \, M}_\odot {\rm
\, yr}^{-1} $). A89B-2 was already excluded as an X-ray AGN due to the expected
X-ray emission from diffuse gas and the large offset between the X-ray and
optical positions; however, HMXBs could account for 30\% of the X-ray emission
from A89B-2. From the combined detections and limits on the SFR from
[\ion{O}{2}], we conclude that unobscured star formation is not likely to be
responsible for the X-ray emission used to identify our X-ray AGN.

Since [\ion{O}{2}] emission can be obscured, one must also consider wavelengths
where obscuration is less of an issue. In the far-infrared (FIR), reradiating
dust reveals obscured star formation. If one considers the far-infrared SFR
relation \citep{K1998}, corrected to the Infrared Astronomy Satellite ({\it
IRAS}) bands \citep{CAB+2000}, the obscured SFR can be estimated from
\begin{equation}
{\rm SFR}_{\rm FIR} \approx 7.9 \times 10^{-44} \,
                            \frac{L_{\rm FIR}}{\rm erg \, s^{-1}}
                            {\rm \, M}_\odot {\rm \, yr}^{-1},
\end{equation}
where $L_{\rm FIR}$ is calculated from the luminosity distance, $D_{L}$, and the 
the {\it IRAS} 60 and 100 $\mu$m fluxes in Janskys,
\begin{equation}
L_{\rm FIR} = 4 \pi D^{2}_{L} \,
              1.26 \times 10 ^{-14} \, (2.58 F_{60} + F_{100}).
\end{equation}
Only two of the X-ray detected galaxies are detected by {\it IRAS}, A85-1 and
A754-6, both X-ray AGN. For A85-1, there are detections at both 60 and 100
$\mu$m, while A754-6 is only detected at 60 $\mu$m. Their predicted SFR$_{\rm
FIR}$ of about 9 and $<16 {\rm \, M}_\odot {\rm \, yr}^{-1}$, respectively imply
approximately 20\% and $<50\%$ of their X-ray emission could come from HMXBs. 
The hard median X-ray energy of A754-6 suggests that whatever source is emitting
X-rays is obscured; an obscured AGN would also reradiate in the FIR. Although
some star formation might be ongoing in these two sources, their identification
as X-ray AGN appears secure.

Given the typical minimum fluxes of detected sources, $F_{60} \sim 0.2 {\rm \,
Jy}$ and $F_{100} \sim 1 {\rm \, Jy}$ in the {\it IRAS} Faint Source Catalog
\citep{M+1990}, rough upper limits to the SFR$_{\rm FIR}$ of about $<11$, $<11$,
$<13$, $<14$, $<18$, and $<22 {\rm \, M}_\odot {\rm \, yr}^{-1}$ can be set for
FIR undetected galaxies in Abell 754, Abell 85, Abell 3128, Abell 3125, Abell
644, and Abell 89B, respectively. Since a SFR of $10 {\rm \, M}_\odot {\rm \,
yr}^{-1}$ could account for X-ray luminosities from HMXBs of $10^{41} {\rm \,
erg \, s}^{-1}$, current SFR limits from {\it IRAS} are too shallow to rule out
a 100\% HMXB origin of the X-ray emission for three of the X-ray AGN, A89B-5,
A644-2, and A3128-9.

In galaxies without a radio AGN, the radio emission at frequencies
below tens of GHz can be a direct probe of the current star formation
of massive stars \citep[$M \ge 5 M\odot$,][]{C1992};
\begin{multline}
{\rm SFR}_{\rm GHz} \approx \frac{L_{\rm GHz}}{{\rm W \, Hz}^{-1}}  \times \\
                             \left[ 5.3 \times 10^{21}
                                      \left(\frac{\nu}{\rm GHz}\right)^{-0.8} +
                                       5.5 \times 10^{20}
                                      \left(\frac{\nu}{\rm GHz}\right)^{-0.1}
                                 \right] ^{-1}
\end{multline}
where $L_{\rm GHz}$ is the radio luminosity measured at frequency $\nu$. The
X-ray detected galaxies that have NVSS counterparts are A85-1 ($7.5 {\rm \,
mJy}$), A89B-3 ($64.6 {\rm \, mJy}$), A754-1 ($158.7 {\rm \, mJy}$), A754-2
($3.2 {\rm \, mJy}$), A754-3 ($7.3 {\rm \, mJy}$), A754-4 ($71.3 {\rm \, mJy}$),
and A754-7 ($81.2 {\rm \, mJy}$). Three of the sources, A754-1, A754-4, and
A754-7, have multiple NVSS components and have been identified as narrow-angle
tail radio AGN \citep{ZBO1989}%
\footnote{Only one narrow-angle tail candidate in Abell 754, PGC 025746, 
is undetected in our X-ray observations \citep{ZBO1989}.}. As such, they are not
suitable for placing limits on the SFR and the extended radio sources
unambiguously reveal the presence of AGN. Since the NVSS counterparts to A89B-3
and A754-2 are offset by $19\farcs4$ and $36\farcs1$ , respectively, they too
are unlikely to be due to star formation. In fact, catalog results from the
higher resolution 1.4 GHz FIRST survey \citep{WBH+1997} indicates that the
A89B-3 is a narrow-angle tail radio AGN, while no FIRST data for A754-2 are
available. Only two of the sources have radio emission that could arise from the
core of a radio jet or galactic star formation, A85-1 and A754-3. If the radio
emission from A85-1 arises from star-formation, its ${\rm SFR}_{\rm GHz} \sim 12
{\rm \, M}_\odot {\rm \, yr}^{-1}$. However, the joint radio and FIR detection
allows calculation of its $q$ parameter
\citep[e.g.,][]{CAH1991,YRC2001}. With $q=1.717\pm0.096$, A85-1 is
well away from the typical value of $q=2.34 $ for star forming galaxies
\citep{YRC2001}, indicative of a radio excess with a likely AGN origin. This
suggests its SFR would be lower than what we measure.
Given its additional classification as a spectroscopic AGN, its identification
as an X-ray AGN seems robust against the effects of star formation. A754-3,
whose detection could be consistent with $11 {\rm \, M}_\odot {\rm \, yr}^{-1}$
of star formation, was not classified as an X-ray AGN due to its low X-ray
luminosity that could be due to LMXBs or diffuse gas.
Since A754-3 is an early-type galaxy \citep[E-S0;][]{PPP+2003} at the center of
Abell 754, the radio emission appears more likely to arise from a low-luminosity
AGN than from star-formation. If star-formation is ongoing, X-ray emission from
HMXBs could also be responsible for the X-ray emission.

With a detection limit of $2.5 {\rm \, mJy}$, the 1.4 GHz NRAO VLA Sky Survey
(NVSS) can place tighter constraints on the current SFR than the FIR. For
sources undetected in the NVSS, upper limits to the SFR of about $<3.9$, $<4.0$,
$<6.5$, and $<7.9 {\rm \, M}_\odot {\rm \, yr}^{-1}$ can be set for galaxies in
Abell 754, Abell 85, Abell 644, and Abell 89B, respectively; Abell 3125 and
Abell 3128 are too far south to have been included in the survey. These limits
are low enough to rule out a strong HMXB origin of the X-ray emission among our
sample of X-ray AGN undetected by NVSS, A85-5 ($<8\%$) , A89B-5 ($<44\%$),
A644-1 ($<5\%$), A644-2 ($<38\%$), and A754-6 ($<10\%$).

From the combined constraints on star formation in our X-ray detected galaxies,
we conclude that star formation is not likely to be responsible for the X-ray
emission used to identify our X-ray AGN. We also conclude that there is not
strong evidence for more than a few highly star-forming galaxies ($ {\rm SFR}
\gtrsim 10 {\rm \, M}_\odot {\rm \, yr}^{-1}$) in these clusters.

\subsection{AGN Fraction and Host Galaxy Magnitude}

\begin{figure*}
\plottwo{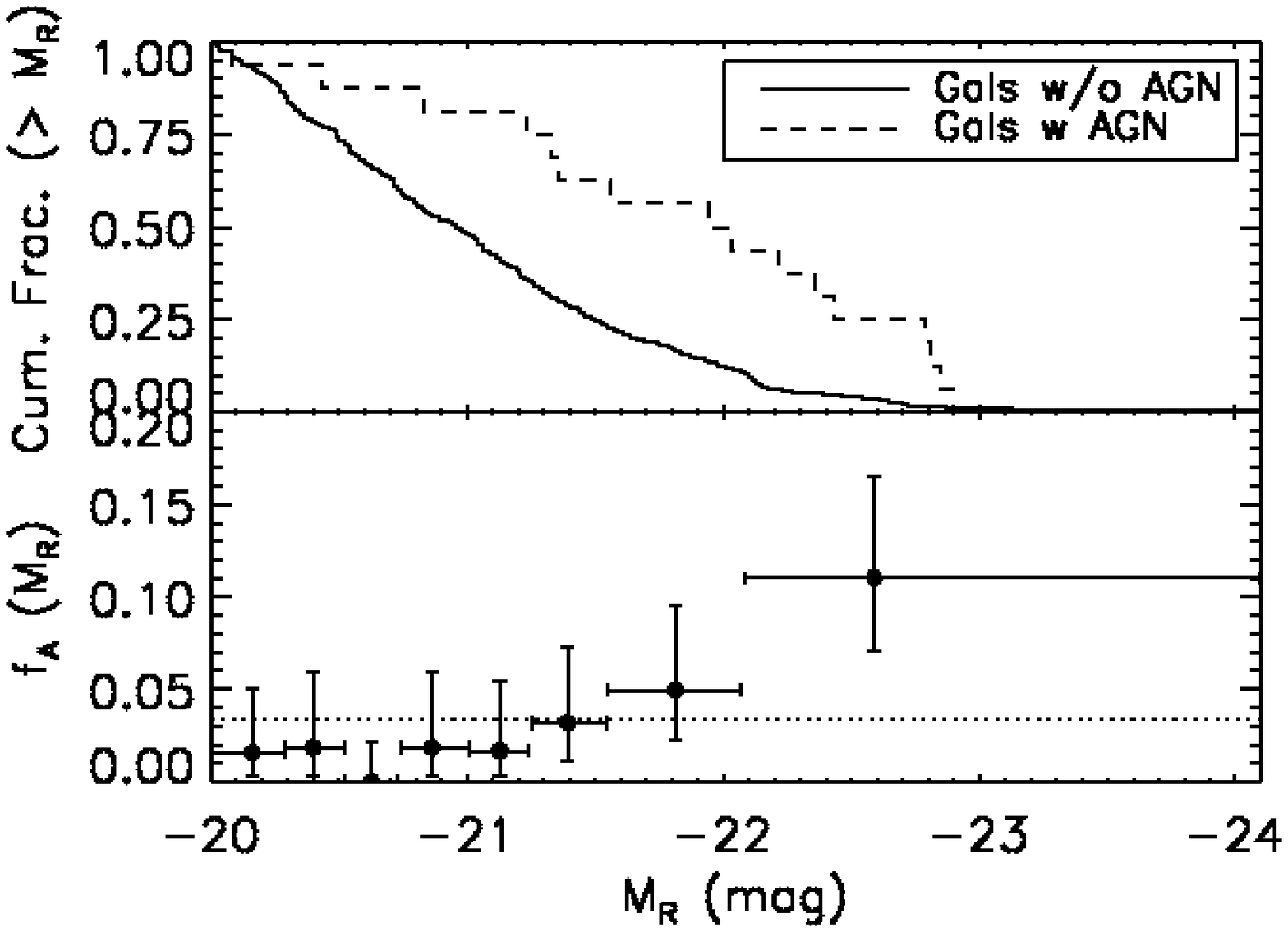}{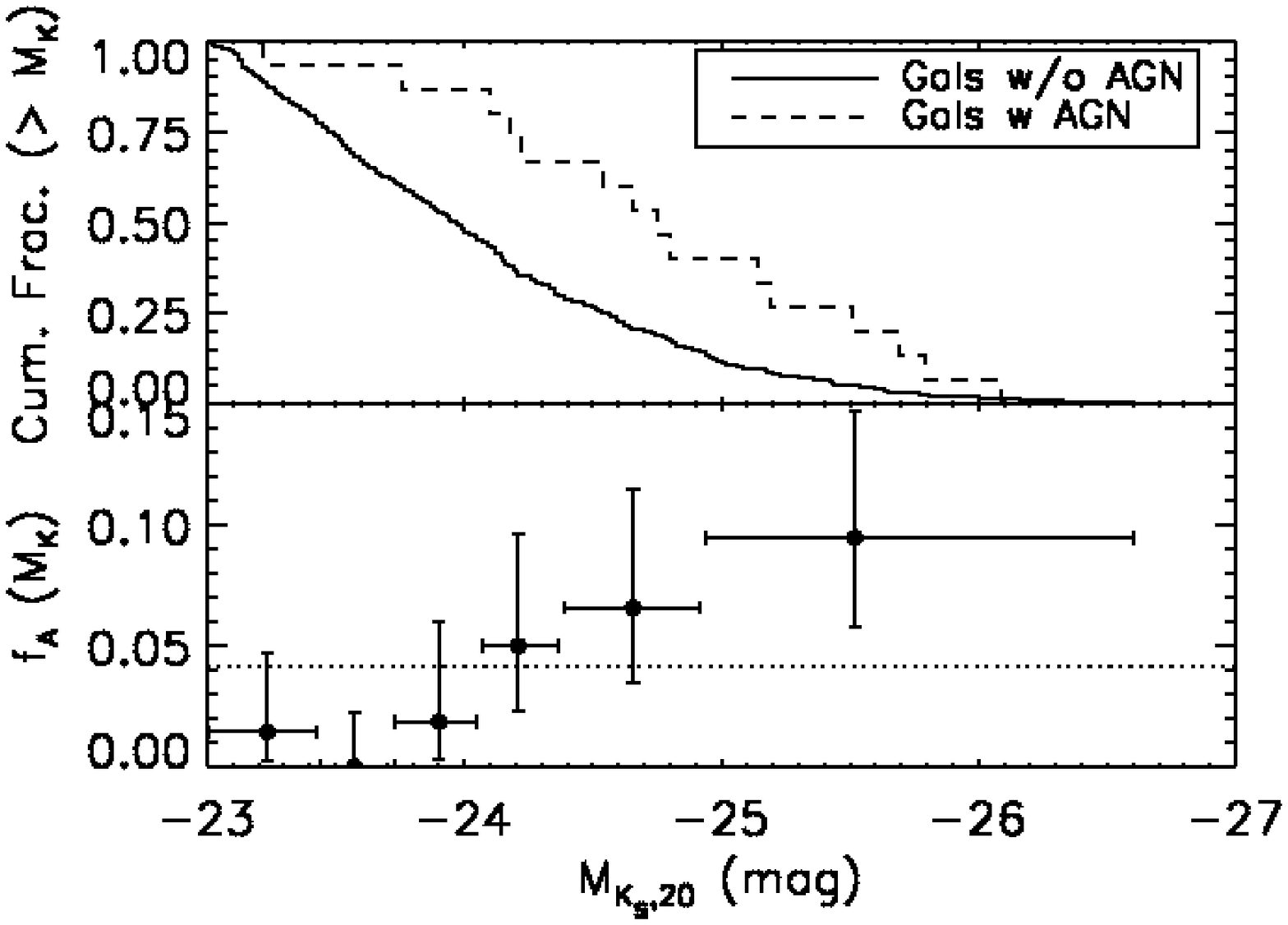}
\caption{
({\it Top}) The cumulative fraction of galaxies with ({\it dashed} line) and
without ({\it solid} line) X-ray AGN as a function of $M_R < -20$ ({\it left}) and
$M_{K_s,K20} < -23.0$ ({\it right}) for galaxies in the cluster sample from
Table~\ref{tab:cluster}. ({\it Bottom}) The fraction of galaxies with an X-ray
AGN, $f_{\rm A}$, in bins of approximately 50 galaxies. 
In both panels, the {\it dotted} line indicates the fraction summing over all $M_R < -20$
galaxies in the sample, $3.4^{+1.1}_{-0.8}$\%.
It is clear that more luminous galaxies are more likely to
contain X-ray AGN.
\label{fig:frac_mag}}
\end{figure*}

For $L_{X,B} > 10^{41} {\rm \, erg \, s}^{-1}$ and $M_R < -20$, the X-ray
identified AGN fraction, $f_{\rm A}(M_R<-20;L_{X,B}>10^{41})$, summing over
galaxies in eight $z\lesssim 0.3$ clusters was 2.2\%, and $f_{\rm
A}(M_R<-21.3;L_{X,B}>10^{41}) = 9.8\%$ \citepalias{MKK+2006}. \citet{SJF+2007}
found $f_{\rm A}(M_R<-21.3;L_{X,B}>10^{41}) \sim 5\%$ for $0.01 < z < 0.05$
clusters.
This presents an indication that the optical luminosity of a host galaxy affects
whether an X-ray AGN is detected at these X-ray luminosities. In determining
these AGN fractions, both samples did not eliminate luminous X-ray galaxies
whose emission could actually be due to diffuse gas. This may affect the
\citet{SJF+2007} X-ray AGN sample more, as half of them were in BCGs, while none
of the \citetalias{MKK+2006} X-ray AGN were. In the following section, we
explore the magnitude dependence of $f_{\rm A}$ for galaxies in our sample,
after applying the $L_X/L_{K_s}$ relations to select X-ray AGN.

We calculated the extinction-corrected, absolute $M_R$ and $M_{K_s,K20}$
rest-frame magnitudes for all available cluster galaxies, as in 
\S~\ref{sec:obs_analysis} and
Table~\ref{tab:ogals}. 
Based on all cluster galaxies with both magnitudes, we find the
distribution of rest-frame colors $M_{R}-M_{K_s}$ is consistent with a Gaussian
distribution centered on 2.82 with a dispersion of 0.26.
Since we were unable to determine if the BCG of Abell 85 contained an X-ray AGN,
we removed it from the sample. In the top panels of Figure~\ref{fig:frac_mag},
we compare the $M_R$ ({\it left}) and $M_{K_s}$ ({\it right}) distributions of
galaxies with AGN ({\it dashed}) and without AGN ({\it solid}).
The
distributions are clearly different; KS tests indicate the probabilities they are
the same are $1.9\times10^{-3}$ and $5.7\times10^{-3}$, respectively. In the
bottom panels of Figure~\ref{fig:frac_mag}, we display $f_{\rm
A}(L_{X,B}>10^{41})$ in magnitude bins. For the galaxies in Abell 644 and Abell
3125, we applied a correction ($\sim 3.9$ and 1.4, respectively; see
Table~\ref{tab:f_a}) to the number of galaxies at a given magnitude to account
for their incomplete membership information \citepalias{MMK2007}.
This was necessary because membership is incomplete at these optical magnitudes,
but any X-ray detected optical source was always targeted for spectroscopy in
\citetalias{MMK2007}.
The spectroscopic measurements for Abell 85, Abell 89B, and Abell 754 are 100\%
complete at $m_R < 16$, but completeness does drop to $\sim 20$--$50\%$ by $m_R
= 18$ \citepalias{CZ2003}. We note that $M_R = -20$ corresponds to $m_r \sim
17.5$ for Abell 89B and $m_r \sim 16.9$ for Abell 85 and Abell 754. Since these
measurements were made prior to our analysis of the X-ray data, the
spectroscopic completeness should be largely independent of the X-ray properties
of galaxies; any completeness correction would equally correct the numerator and
denominator in the fractions involved. Although one might be concerned that AGN
with emission lines are more likely to have a measured redshift, we note that
only a small fraction, $\sim 20\%$ of X-ray AGN have such emission lines. Any
correction for such an effect would be smaller than the current error bars on
AGN fractions, which are limited by the small numbers of AGN. We further note
that we found no X-ray source matched to a photometric object without a redshift
that would be consistent with $M_R < -20$.

Since it is difficult to construct a clear situation where the X-ray
completeness depends on the optical/near-IR host magnitude of the galaxy, we do not
expect that sources not detected as X-ray AGN due to X-ray
incompleteness are the cause of this discrepancy. To explicitly test this, we
considered two sets of X-ray AGN at brighter luminosities, X-ray AGN with
$L_{X,B} > 4\times 10^{41} {\rm \, erg \, s}^{-1}$ over the entire sample, and
X-ray AGN with $L_{X,B} > 2\times 10^{41} {\rm \, erg \, s}^{-1}$ when removing
all A89B galaxies from consideration. In both case, we still find that the
optical/near-IR magnitudes of galaxies with and without X-ray AGN are drawn from
different distributions using the KS test.

\begin{deluxetable*}{lrrcrrcrr}
\tabletypesize{\footnotesize}
\tablewidth{0pt}
\tablecaption{AGN Fraction 
\label{tab:f_a}}
\tablehead{
\colhead{Cluster} &
\multicolumn{2}{c}{--- $N_{\rm XAGN} (M_R<-20)$ ---}&
&
\multicolumn{2}{c}{--- Members $(M_R<-20)$ --- }&
&
\multicolumn{2}{c}{--------- $f_{\rm A}$ --------- }\\
\colhead{Name} &
\colhead{$L_{X,B}>10^{41}$} &
\colhead{$L_{X,B}>10^{42}$} &
&
\colhead{Confirmed} &
\colhead{Corrected} &
&
\colhead{$L_{X,B}>10^{41}$} &
\colhead{$L_{X,B}>10^{42}$}\\
\colhead{(1)} &
\colhead{(2)} &
\colhead{(3)} &
&
\colhead{(4)} &
\colhead{(5)} &
&
\colhead{(6)} &
\colhead{(7)}
}
\startdata
Abell 85   &  2 & 0 & & 109 & 109 & & $0.018^{+0.024}_{-0.012}$ & $0.000^{+0.010}_{-0.000}$ \\
Abell 89B  &  2 & 1 & &  22 &  22 & & $0.091^{+0.108}_{-0.058}$ & $0.045^{+0.097}_{-0.038}$ \\
Abell 3125 &  3 & 0 & &  20 &  28 & & $0.107^{+0.093}_{-0.058}$ & $0.000^{+0.040}_{-0.000}$ \\
Abell 3128 &  4 & 1 & &  67 &  67 & & $0.060^{+0.045}_{-0.028}$ & $0.015^{+0.033}_{-0.012}$ \\
Abell 754  &  3 & 1 & & 171 & 171 & & $0.018^{+0.017}_{-0.010}$ & $0.006^{+0.013}_{-0.005}$ \\
Abell 644  &  2 & 1 & &  19 &  75 & & $0.027^{+0.034}_{-0.017}$ & $0.013^{+0.030}_{-0.011}$ \\
\\
Average    &    &   & &     &     & & $0.031^{+0.011}_{-0.011}$ & $0.009^{+0.006}_{-0.006}$ \\
Sum        & 17 & 4 & & 408 & 472 & & $0.034^{+0.011}_{-0.008}$ & $0.008^{+0.007}_{-0.004}$ \\

\enddata
\tablecomments{AGN fractions for $M_R < -20$ galaxies in six $z<0.08$ clusters.
Columns are:
(1) Cluster Name;
(2) Number of X-ray AGN with $L_{X,B} > 10^{41} {\rm \, erg \, s}^{-1}$;
(3) Number of X-ray AGN with $L_{X,B} > 10^{42} {\rm \, erg \, s}^{-1}$;
(4) Number of $M_R < -20$ galaxies with spectroscopically confirmed redshifts within {\it Chandra} FOV;
(5) Number of $M_R < -20$ galaxies within {\it Chandra} FOV, corrected for
preferential spectroscopic targetting of X-ray detections;
(6) X-ray AGN fraction with $L_{X,B} > 10^{41} {\rm \, erg \, s}^{-1}$;
(7) X-ray AGN fraction with $L_{X,B} > 10^{42} {\rm \, erg \, s}^{-1}$;
}
\end{deluxetable*}

Since the X-ray emission from LMXBs and diffuse gas also increases with optical/near-IR
magnitude, one concern is that the higher fractions of X-ray AGN at brighter
magnitudes is due to normal X-ray emitting galaxies that are misidentified as
X-ray AGN. To address this, we have performed Monte Carlo simulations to
determine how severely our AGN sample could be contaminated by the combination of
X-ray emission from LMXBs and diffuse gas.

\begin{figure}
\plotone{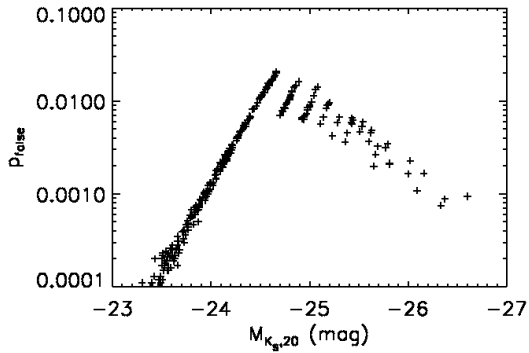}
\caption{
Probability a cluster galaxy would be falsely identified as an X-ray AGN
as a function of $M_{K_sK20}$. For the vast majority of cluster galaxies
the probability that X-ray emission from LMXBs and diffuse gas would be
misclassified as an X-ray AGN is small. The total number of falsely
identified AGN over the entire sample is $<1.1$.
\label{fig:false}}
\end{figure}

We performed $10^5$ realizations of the X-ray to near-IR luminosity relations for
both LMXBs (eq.\ \ref {eq:l_x_lmxbs}) and diffuse gas (eq.\ \ref {eq:l_x_gas}, with
proper corrections), assuming that the reported errors in the relations
follow Gaussian statistics. Within each realization, we combined the two simulated
relations to predict the X-ray luminosity from LMXBs and diffuse gas for each cluster
galaxy. Since our measured luminosities were for point sources, while
the predicted luminosities were for entire galaxies, we needed to determine and
apply a correction factor. Comparisons of the point-source counts and the counts
within the $K_s = 20 {\rm \, mag \, arcsec}^2$ isophote for the X-ray detected
galaxies not classified as X-ray AGN indicated that the average luminosity of
the entire galaxy was 1.5 times that for a point source. We applied this
correction to the predicted luminosities and then converted to an expected
number of counts for each realization assuming an exposure time appropriate for
the flanking field observations.
This expected number of counts was used to create a simulated output number of
counts assuming random deviations drawn from Poisson distributions with the
expected number of counts as its mean. The output number of counts and its
error were then converted back into luminosity, $L_{X,B,MC}$, and we performed
the same (luminosity) selection criteria to identify a source as an X-ray AGN:
$L_{X,B,MC} > 10^{41} {\rm \, erg \, s}^{-1}$ and $L_{X,B,MC}$ more than
$1\sigma$ higher than the sum of the upper limits to the \citet{KF2004} and
\citet{SJF+2007} relations. Thus, for each cluster galaxy we could calculate the
probability that a source was misidentified as an X-ray AGN, $p_{\rm false}$
(Figure~\ref{fig:false}).

For $M_{K_s} \gtrsim -24.8$,
$p_{\rm false} \lesssim 0.02$ and drops as host galaxy optical luminosity
decreases due to the $10^{41} {\rm \, erg \, s}^{-1}$ minimum X-ray luminosity
for X-ray AGN classification. For $M_{K_s} \lesssim -25.2$, $p_{\rm false}
\lesssim 0.01$, and roughly drops as host galaxy optical luminosity increases.
This is due to the X-ray AGN selection criteria set by the X-ray emission
expected from LMXBs and gas. The roughly diagonal line up to $M_{K_s} \gtrsim
-24.8$ corresponds to a minimum of 5 counts. Similar rough diagonal lines at
increasing galaxy optical luminosity corresponds to an increasing minimum number
of counts that satisfy our criteria. As the exposure times are made larger, as
in our central field observations, the values of $p_{\rm false}$ drop at a given
$M_{K_s}$. Therefore, calculations based on Figure~\ref{fig:false}
represent the most conservative, i.e., highest, estimate of the number of
sources in our entire sample that are falsely identified as X-ray AGN due to
their LMXB and diffuse gas emission. By summing the probabilities, we estimate
that there are $<1.1$ such false sources.
Approximately 0.3, 0.6, and 0.2 falsely indentified sources are predicted for
the three brightest bins, respectively, in each of the bottom panels of
Figure~\ref{fig:frac_mag}. This would reduce their fractions by $\lesssim
0.005$, 0.010, and 0.003, which is much less than the current error bars.
In addition, we note that even after removing the three AGN closest to the
expected relation combining LMXBs and diffuse gas, the KS test still indicates
that the $M_R$ ({\it left}) and $M_{K_s}$ ({\it right}) distributions of
galaxies with and without AGN are not drawn from the same distributions. We
conclude that misidentified X-ray emission from LMXBs and diffuse gas are not
responsible for X-ray detected AGN being more likely to be found in more
luminous galaxies.

\begin{figure}
\plotone{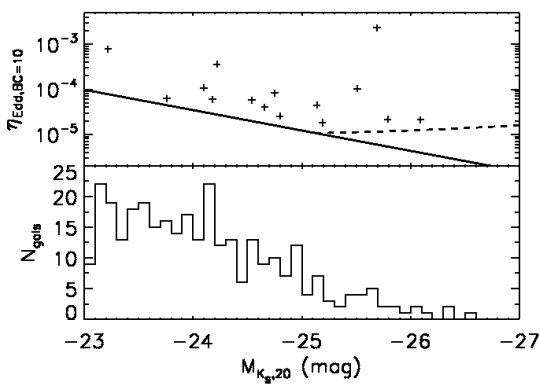}
\caption{
({\it Top}) Approximate Eddington-normalized accretion rates, $\eta_{Edd}$,
of detected X-ray AGN as a function of $M_{K_sK20}$. We assume
a bolometric correction to the X-ray luminosity, ${\rm BC}=10$. Since
we assumed all  galaxies are spheroids, their black hole mass may be lower
and the normalized accretion rates may be higher at a given $M_{K_sK20}$.
The {\it solid} line indicates the limit set by 
$L_{X,B} > 10^{41} {\rm \, erg \, s}^{-1}$.
The {\it dashed} line indicates the limit set to exclude potential contributions
from LMXBs and diffuse gas.
The changing AGN fraction with host
galaxy luminosity, Figure~\ref{fig:frac_mag}, is likely due to a selection bias
where lower Eddington accretion efficient sources are only detectable in
more luminous galaxies, which are more likely to have more
massive black holes. ({\it Bottom}) The histogram of $M_{K_sK20}$ cluster
member galaxies is shown for reference.
\label{fig:eff}}
\end{figure}

Our results that X-ray detected AGN are more likely to be found in more luminous
galaxies
for $M_R$
are consistent with results from the XMM detections of AGN in
the Abell 901/902 supercluster \citep[$z\sim0.17$][]{GGO+2007}. Both show a
nearly constant fraction of X-ray AGN of $\sim1\%$ for $-21.5 \lesssim M_R
\lesssim -20$ galaxies. For brighter galaxies, this increases to a maximum of
$\sim 10$--$20\%$. We can use $M_{K_{s},K20}$ to derive galactic stellar masses
for these galaxies. 
If we use
\citet{BC2003}, assuming their Padova 1994 evolutionary tracks for a $13 {\rm \,
Gyr}$, solar abundance single stellar population with the
\citet{C2003} initial mass function, and correct the isophotal magnitude to a
total magnitude, the magnitudes in Figure~\ref{fig:frac_mag} ({\it right})
correspond to about $5 \times 10^{10}$ -- $2 \times 10^{12} {\rm \, M}_\odot$. A
similar trend has been observed for radio-loud AGN, where the fraction of
radio-loud AGN increases with stellar mass, $f \propto M_*^{2.5}$ up to $\sim
10^{12} {\rm \, M}_\odot$ \citep{BKH+2005}.
On the other hand, the fraction of strong ($L[\mion{O}{3}] > 10^7 {\rm \,
L}_\odot)$, optically identified AGN drops from $\sim 12\%$ at $10^{11} {\rm \,
M}_\odot$ to $\sim 2\%$ at $10^{10} {\rm \, M}_\odot$ and $10^{12} {\rm \,
M}_\odot$ \citep{KHT+2003}. The physically relevant question that these trends
raise is whether AGN are more likely to reside in more luminous (massive) host
galaxies.

In the case of our X-ray AGN, much of the signal may actually be due to a
physical selection bias.
For each of our X-ray AGN, we
can roughly estimate the accretion rate relative to the Eddington value, divided
by the bolometric correction , $\eta_{\rm Edd} / {\rm BC}$, from $L_{X,B}$.
As near-IR light is a better tracer of mass and subject to less extinction than
visible light, we converted the detected near-IR luminosity to a black hole mass
\citep{MH2003} under the assumption that all the luminosity is from a bulge
component, $\log M_{\rm BH} = 8.21+1.13 (\log L_{K_s,{\rm ttl}})$.
Since these sources are likely to have a range of disk contributions
to their luminosities
our derived $\eta_{\rm Edd} / {\rm BC}$ are underestimated; however, our
results are still illustrative. Although BC is uncertain, especially if the
mechanism for low and high luminosity X-ray AGN differ, we assume ${\rm BC} =
10$ and display, $\eta_{{\rm Edd,BC}=10}$ in Figure~\ref{fig:eff}.
This assumption is reasonable given derived values of BC at
these X-ray luminosities \citep[in the $2$--$10 {\rm \, keV}$ band;][]{VF2007}.
The detected
sources cover ranges of $1\times 10^{-5}
\lesssim \eta_{{\rm Edd,BC}=10} \lesssim 3\times 10^{-3}$. Given our requirement
that $L_{X,B} > 10^{41} {\rm \, erg \, s}^{-1}$, we can calculate the minimum
$\eta_{{\rm Edd,BC}=10}$.
It is clear that at higher luminosities, X-ray AGN at
lower $\eta_{{\rm Edd,BC}=10}$ can, and are, detected.
Similar results are also
seen in Abell 901/902 \citep{GGO+2007}; their $\eta_{\rm
Edd}$ are $\sim5$ higher because they adopted the
$R$-band host-galaxy luminosity black hole mass
relation of \cite{MD2002}.
Calculated AGN fractions will only be independent of the galaxy
luminosity function when they are sensitive to the same $\eta_{\rm Edd}$.

\begin{figure*}
\plottwo{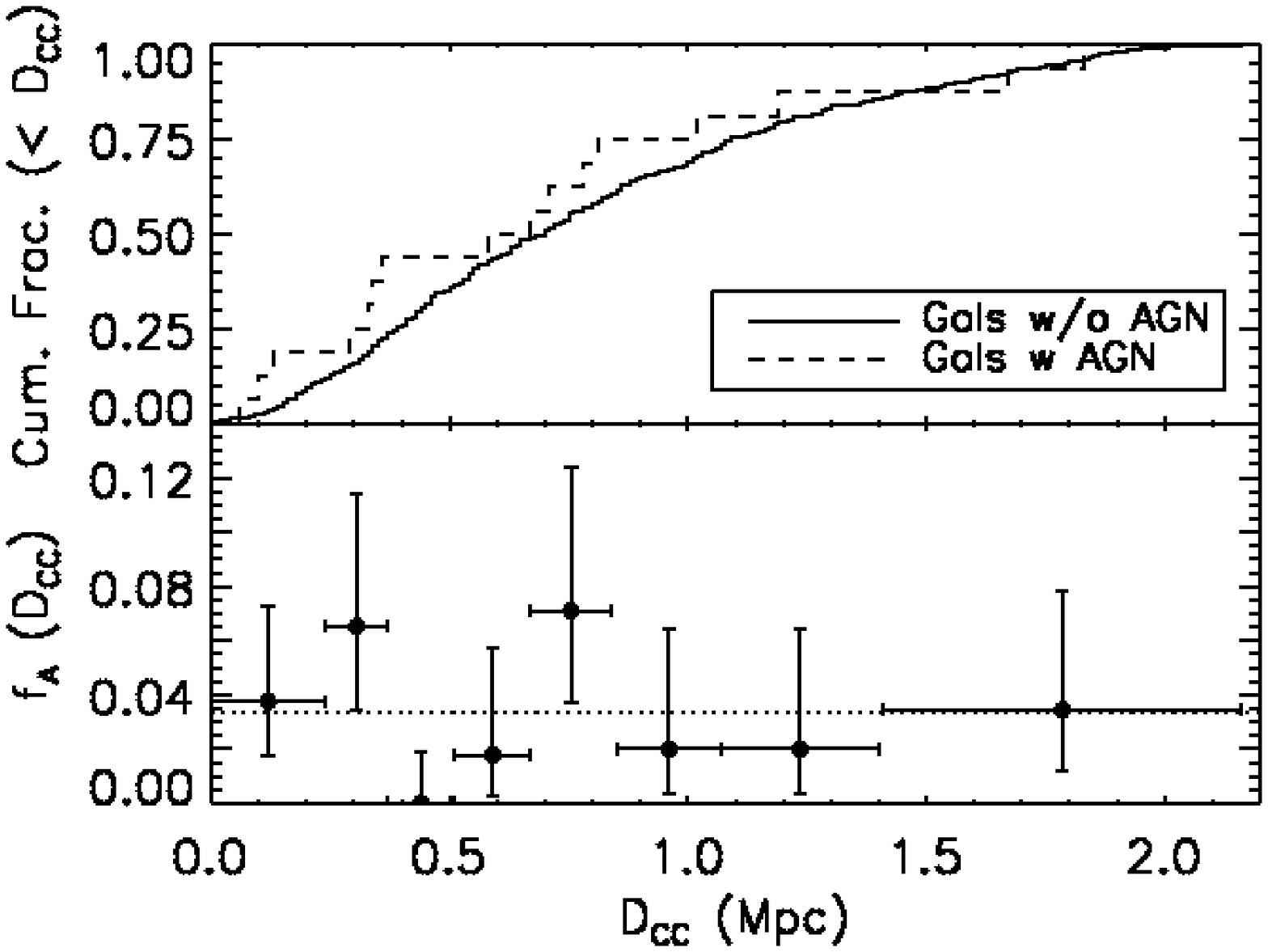}{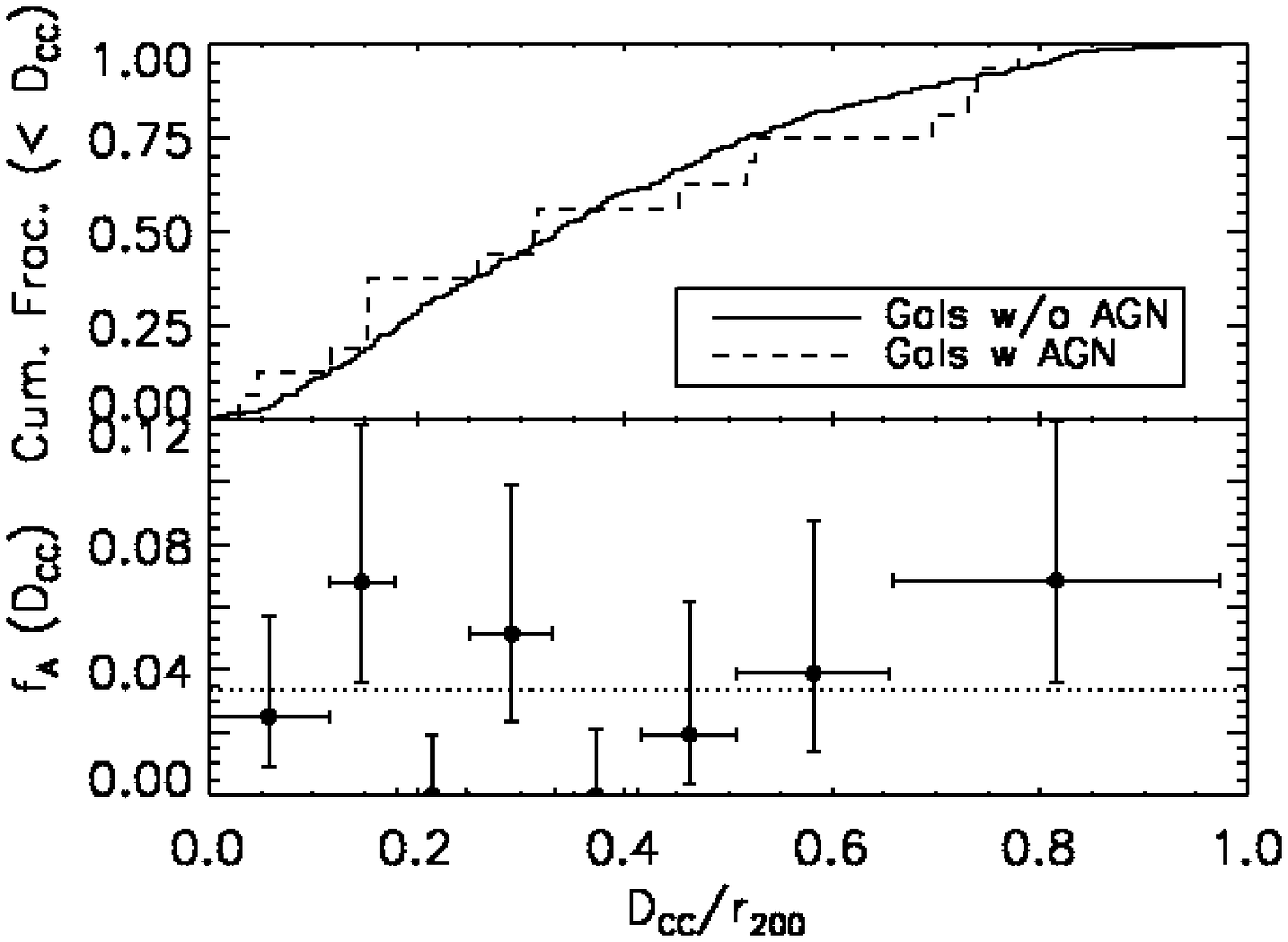}
\caption{
({\it Top}) The cumulative fraction of $M_R < -20$ galaxies with ({\it dashed}
line) and without ({\it solid} line) X-ray AGN as a function of cluster-centric
distance, $D_{CC}$, in Mpc ({\it left}) and units of $r_{200}$ ({\it right}) for
galaxies in the cluster sample from Table~\ref{tab:cluster}. ({\it Bottom}) The
fraction of galaxies with an X-ray AGN, $f_{\rm A}$, in bins of approximately 50
galaxies. The {\it dotted} line indicates the fraction for the $M_R < -20$
sample.
The radial distributions of galaxies with and without an X-ray AGN are
comparable, consistent with \citetalias{MMK2007} results for $L_{X,B} > 10^{41}
{\rm \, erg \, s}^{-1}$ and \citet{RE2005} results for disturbed clusters.
\label{fig:frac_dist}}
\end{figure*}

Since more massive galaxies tend to be more centrally concentrated
\citep[e.g.,][]{KHW+2003}, indicative of a more dominant bulge component,
our $\eta_{{\rm Edd,BC}=10}$ are likely to represent larger underestimates at
lower near-IR luminosities than at higher luminosities. Thus, the disparity
between the $\eta_{{\rm Edd,BC}=10}$ probed by a given X-ray luminosity 
at lower and higher near-IR luminosities is even larger than that suggested by
Figure~\ref{fig:eff}.

There are two implication of this selection bias. First, this bias
makes it essential that comparisons of AGN fractions make the same assumptions
in both their X-ray luminosity and optical magnitude cuts,
and that host galaxy morphology (spheroid mass) may also be important.
More importantly, 
$\eta_{\rm Edd} / {\rm BC}$ is a more physical measure of AGN activity than
the X-ray luminosity. The true fraction of $M_R < -20$ galaxies that host
X-ray AGN with $\eta_{\rm Edd} / {\rm BC} \gtrsim 10^{-6}$ will be larger than
the $\sim3\%$ we measure.

\subsection{Radial Distribution of AGN}

One of the goals of the new observations of Abell 85 and Abell 754 was to
measure the spatial distribution of X-ray detected AGN. Along with Abell 89B,
these three clusters have partial coverage out to $\sim r_{200}$. Although the
radial coverage of the other three clusters is smaller, we include them in our
measurement of the radial distribution. The {\it Chandra} observations of Abell
644 and Abell 3128 have coverage out to $\sim 0.4 r_{200}$, while the one of
Abell 3125 extends farther to $\sim 0.7 r_{200}$.
Since we never include
galaxies outside of the {\it Chandra} FOVs, including all six clusters
will improve the statistical determination of the AGN fraction with radius,
particularly for smaller radii.

In the top panels of Figure~\ref{fig:frac_dist}, we compare the (projected)
spatial distributions of $M_R < -20$ galaxies with AGN ({\it dashed}) and
without AGN ({\it solid}). We consider the distributions as a function of
physical distance ({\it left}) and distance scaled to $r_{200}$ ({\it right}). 
In the bottom panels of Figure~\ref{fig:frac_dist}, we display
$f_{\rm A}(M_R<-20;L_{X,B}>10^{41})$ in spatial bins, correcting for membership
completeness. For this sample of galaxies, the spatial distributions of galaxies
with X-ray AGN are the same as galaxies without X-ray AGN.

As X-ray incompleteness is only a minor issue at the edges of the flanking field
observations of Abell 85 and Abell 754 and the single observation of Abell 3125,
we do not believe it is masking an increased contribution of AGN at
the outskirts of clusters. The upper error bars presented
in any single bin in the bottom panels of Figure~\ref{fig:frac_dist} correspond
to missing $\sim1$--3 sources in a given bin. Since the edges of the flanking
fields cover a wide range of cluster-centric distances, any missing sources should
be spread over multiple bins. Thus, the effect from missing only one or two sources
over the entire sample is well within our current noise level.
In addition, we do not find any radial
dependence in the smaller samples of AGN where we are complete; $L_{X,B} >
4\times 10^{41} {\rm \, erg \, s}^{-1}$ in the entire sample or $L_{X,B} >
2\times 10^{41} {\rm \, erg \, s}^{-1}$ when removing all A89B galaxies from
consideration.

In \citetalias{MMK2007}, the spatial distribution of
$L_{X,B} > 10^{41} {\rm \, erg \, s}^{-1}$ X-ray AGN was also consistent
with the other cluster members. We do note that our and their distribution are not
independent as they share Abell 644, Abell 3125, and 3128. There are too few X-ray AGN with
$L_{X,B} > 10^{42} {\rm \, erg \, s}^{-1}$ in our sample to test their
result that the more X-ray luminous AGN are more centrally concentrated.

\begin{figure}
\plotone{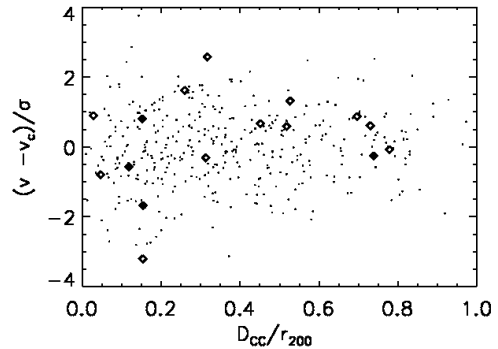}
\caption{
Radial velocities of cluster $M_R < -20$ galaxies relative to the mean velocity
of each cluster and normalized by the cluster velocity dispersion as a function
of cluster-centric distance, $D_{CC}$, in units of $r_{200}$. Diamonds indicate
galaxies detected as X-ray AGN. Filled diamonds indicate $L_{X,B} > 10^{42} {\rm
\, erg \, s}^{-1}$. The two-dimensional distributions of galaxies with and
without an X-ray AGN are comparable.
\label{fig:vel_pos}}
\end{figure}

One concern is that galaxies with a small projected distance may actually be
infalling galaxies at large physical distances close to the line-of-sight to the
cluster. In Figure~\ref{fig:vel_pos}, we plot the infall velocity relative to
the mean cluster velocity scaled by the cluster velocity dispersion against the
projected distance scaled to $r_{200}$ for $M_R < -20$ galaxies with AGN ({\it
circles}) and without AGN ({\it diamonds}). Although there are two AGN with
large infall velocities at small projected distances, A85-1 and A644-2, a
two-dimensional KS test \citep{PTV+1992} indicates that the galaxies with and
without AGN are not likely to be drawn from samples with different two
dimensional distributions.

If AGN are fueled by galaxy-galaxy interactions, one expects AGN should
be more prevalent in the outskirts of clusters. However, a significant fraction
of early type galaxies, which tend to lie in the centers of richest clusters,
are known to harbor low-luminosity AGN and LINERs. A relation between AGN and
early-type galaxies could dilute or even reverse the trends predicted by
gas-rich mergers or galaxy harassment. In addition, our detected X-ray AGN are
detected more often in more luminous host galaxies. Since more massive galaxies
tend to be early-type galaxies, any observed radial trend due to low-luminosity
AGN and LINERs should be larger than it is intrinsically. The combination of all
three effects could explain why the radial distribution of X-ray AGN is not
markedly different from that of all galaxies in our sample. Another potential
explanation is that radial-averaging over the substructure in the cluster
masks any underlying signal. Finally, there could actually be no preferred
radial distribution for X-ray AGN.

\section{AGN Fraction and Cluster Properties}
\label{sec:cluster}

\citetalias{MMK2007} found some evidence that the AGN fraction varies with the
properties of the host cluster. When summing over galaxies, their $f_{\rm
A}(M_R<-20;L_{X,B}>10^{41})$ is 2.2\%; however their cluster-averaged $f_{\rm
A}(M_R<-20;L_{X,B}>10^{41})$ is 4.9\%. Compared to the cluster-by-cluster
$f_{\rm A}(M_R<-20;L_{X,B}>10^{41})$, a $\chi^2 = 17.8$ for 7 degrees of freedom
(dof) indicates that cluster variations of the AGN fraction are significant. 
They find weak evidence that the AGN fraction is higher at lower redshift, in
lower velocity-dispersion clusters, in clusters with substantial substructure,
and in clusters with a smaller fraction of Butcher-Oemler galaxies. However,
they caution that correlations between several of these parameters
preclude identification of which cluster property or properties most strongly
influence the AGN fraction.

In Table~\ref{tab:f_a} we summarize the AGN fractions for each cluster in our
sample. We first list the number of X-ray AGN more luminous than $10^{41} {\rm
\, erg \, s}^{-1}$ (column 2) and $10^{42} {\rm \, erg \, s}^{-1}$ (column 3). 
We then list the number of spectroscopically confirmed, $M_R < -20$ members in
each cluster (column 4). Since some of the galaxies within \citetalias{MKK+2006}
clusters were preferentially targeted for spectroscopy based on X-ray
detections, and completeness for membership of two of the clusters (Abell 644
and Abell 3125) is not 100\%, we also list a corrected number of $M_R < -20$
galaxies members in each cluster (column 5). We use these numbers to calculate
the X-ray AGN fraction above $10^{41} {\rm \, erg \, s}^{-1}$ (column 6) and
$10^{42} {\rm \, erg \, s}^{-1}$ (column 7). As in \citetalias{MMK2007}, we also
calculate the cluster-by-cluster average AGN fractions, and the AGN fraction
assuming it is independent of cluster properties. We find that the AGN fractions
summing over galaxies, $f_{\rm A}(M_R<-20;L_{X,B}>10^{41})= 3.1^{+1.1}_{-1.1}\%$
and $f_{\rm A}(M_R<-20;L_{X,B}>10^{42})= 0.9^{+0.6}_{-0.6}\%$, are consistent
with the cluster-averaged AGN fractions $f_{\rm A}(M_R<-20;L_{X,B}>10^{41})=
3.4^{+1.1}_{-0.8}\%$ and $f_{\rm A}(M_R<-20;L_{X,B}>10^{42})=
0.8^{+0.7}_{-0.4}\%$. Comparing $f_{\rm A}(M_R<-20;L_{X,B}>10^{41})$
of each cluster to the cluster-averaged AGN fraction, the $\chi^2 = 4.8$ for 5
dof. We note that these numbers have not been corrected for
the one or two sources that may be missed due to X-ray incompleteness. This
change is minimal as the upper error bars due to the small number of X-ray AGN
correspond to missing $\sim2$--3 sources for any given cluster, and $\sim5$
sources for the entire sample.

With our current data for Abell 85, Abell 89, and Abell 754, we choose to
concentrate on the relation between AGN fraction and two cluster properties,
redshift and velocity dispersion, in the paragraphs below. With the narrow
redshift range, but wide velocity dispersion range, we have a greater ability to
break the degeneracy between the two that was present in \citetalias{MMK2007}.

\subsection{AGN Fraction and Redshift}
\label{sec:cluster_z}

Since our sample spans a narrow redshift range, we consider other samples to test
whether there is any redshift evolution of the X-ray AGN fraction. To compare
the AGN fractions we determine to those in \citetalias{MMK2007}, the overlapping
clusters, Abell 644, Abell 3125, and Abell 3128 must be removed from
\citetalias{MMK2007}. The remaining five clusters form a $0.15 < z < 0.32$
sample. In this sample, the AGN fractions summing over galaxies, $f_{\rm
A}(M_R<-20;L_{X,B}>10^{41})= 1.4^{+0.4}_{-0.3}\%$ and $f_{\rm
A}(M_R<-20;L_{X,B}>10^{42})= 0.8^{+0.4}_{-0.3}\%$, are consistent with the
cluster-averaged AGN fraction $f_{\rm A}(M_R<-20;L_{X,B}>10^{41})=
1.2^{+0.5}_{-0.5}\%$ and $f_{\rm A}(M_R<-20;L_{X,B}>10^{42})=
0.9^{+0.4}_{-0.4}\%$. One might then conclude that the AGN fraction at lower
redshifts is indeed higher based on $f_{\rm A}(M_R<-20;L_{X,B}>10^{41})$, as
suggested in \citetalias{MMK2007}; however, \citetalias{MMK2007} also note that
most of the higher redshift sample is not X-ray complete down to $10^{41} {\rm
\, erg \, s}^{-1}$. At $L_{X,B} >10^{42} {\rm \, erg \, s}^{-1}$, where the
sample should be complete, there is no evidence for redshift evolution in
$f_{\rm A}(M_R<-20;L_{X,B}>10^{42})$. We also compare our AGN fraction to that
of the $0.01 < z < 0.05$ sample of \citet{SJF+2007}. We measure $f_{\rm
A}(M_R<-21.3;L_{X,B}>10^{41}) = 6.7^{+2.4}_{-1.9}\%$, while \citet{SJF+2007}
find $f_{\rm A}(M_R<-21.3;L_{X,B}>10^{41}) = 5.5^{+2.4}_{-1.8}\%$. Thus, we
believe that there is no measurable redshift evolution in the X-ray AGN fraction
for $z\lesssim 0.3$ in the current samples. As the \citet{SJF+2007} sample is of
more nearby clusters, it samples a more centrally concentrated population of
galaxies than the sample in this paper. Due to the combination of our errors
being limited by small numbers of AGN and our result that there is no preferred
radial distribution of AGN, this mismatch is not likely to play a large role in
this conclusion.

Recently, a large fraction of luminous X-ray AGN in $z \sim 0.6$ clusters was
measured \citep{EMS+2007}. Due to the redshift of these clusters and the
sensitivity of the observations, fractions were measured for hard
($2.0$--$10.0$) band X-ray luminosities, $L_{X,H}$ above $10^{42} {\rm \, and \,
} 10^{43} {\rm \, erg \, s}^{-1}$. They find $f_{\rm A}(M_R<-20;L_{X,H}>10^{42})
= 2.8^{+1.5}_{-1.0}\%$ $f_{\rm A}(M_R<-20;L_{X,H}>10^{43}) =
2.0^{+1.0}_{-0.7}\%$, and that these fractions were $\sim20$ times that of
$z\sim0.2$ clusters, which is much larger than the expected increases, factors of
1.5 and 3.3, from the measured evolution of the field AGN space density
\citep{UAO+2003}. The largest statistical uncertainty came from the lower
redshift sample. Although we note that there are issues left to explore
regarding the evolution of the AGN fraction in clusters (e.g., the $z \sim 0.6$
clusters are not necessarily the progenitors of the $z \sim 0.2$ clusters;
$M^*_R$ is $\sim 0.4$ brighter at $z\sim0.6$ than at $z\sim0.2$), we can add the
results of Abell 85, Abell 89B, and Abell 754 to \citet{MMK2007} to refine the
estimate for $z\sim0.2$ clusters. In the hard-band, A754-1 is $\sim 9.9
\times 10^{42} {\rm \, erg \, s}^{-1}$ before pileup corrections. These
corrections are likely to make it more luminous than $10^{43} {\rm \, erg \,
s}^{-1}$. No other AGN in these clusters has $L_{X,H}>10^{42} {\rm \, erg \,
s}^{-1}$. Thus for $z \sim 0.2$ clusters, we find $f_{\rm
A}(M_R<-20;L_{X,H}>10^{42}) = 0.18^{+0.17}_{-0.10}\%$ and $f_{\rm
A}(M_R<-20;L_{X,H}>10^{43}) = 0.12^{+0.16}_{-0.08}\%$. These fractions are
consistent with the fractions reported in \citet{EMS+2007}, but with
smaller confidence intervals.
We note that the increased spatial coverage provided by the clusters in this
paper also provides a better match to the more distant clusters, which are
sampled out to their projected $r_{200}$.

\subsection{AGN Fraction and Velocity Dispersion}

The lack of a radial dependence of X-ray AGN fraction in our sample may be
due to the true absence of a trend or the masking of the expected increasing
trend with radius by several other factors, including our increased sensitivity
to AGN in massive galaxies, which tend to lie in cluster cores, and the
significant population of known low-luminosity AGN/LINERs in early-type
galaxies, which are also more numerous in cluster cores. Any trend might also
have been diluted by our averaging over any substructures at a given radius. 
With better AGN statistics, it may be possible to consider whether the AGN
fraction increases in group-like substructures in the cluster relative to the
cluster core, a truer test of the hypothesis that mergers drive AGN today. For
now, we employ another test of the effect of environment on AGN fraction and
thus of the merger scenario: is there a change in X-ray AGN fraction as the
velocity dispersions of clusters increase?

\begin{figure}
\plotone{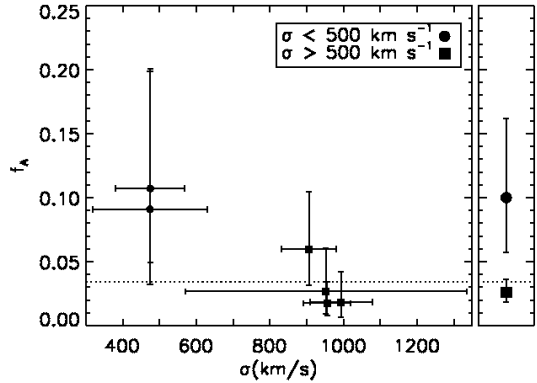}
\caption{
({\it Left}) The fraction of $M_R < -20$ galaxies with X-ray AGN, $f_{\rm A}$,
versus cluster velocity dispersion in the cluster sample from
Table~\ref{tab:cluster}. The {\it dotted} line indicates the fraction summing
over all $M_R < -20$ galaxies in the sample.
({\it Right}) Values of $f_{\rm A}$
summing over galaxies in clusters with velocity dispersions under and over $500
{\rm \, km s}^{-1}$ are indicated with large symbols. The clusters with the
lowest velocity dispersion have the highest fractions, consistent with a
preference for X-ray AGN in the regions with a group-like environments from the
Abell 901/902 supercluster \citep{GGO+2007}.
\label{fig:frac_cl}}
\end{figure}

We display $f_{\rm A}(M_R<-20;L_{X,B}>10^{41})$ as a function of cluster
velocity-dispersion for our $z \lesssim 0.08$ cluster sample in
Figure~\ref{fig:frac_cl}. In our sample, comparing the AGN fraction of each
cluster to the cluster-averaged AGN fraction does not indicate a strong
variation. However, we find a correlation between AGN fraction and velocity
dispersion. Clusters with lower velocity dispersion have larger $f_{\rm
A}(M_R<-20;L_{X,B}>10^{41})$ in our data. In particular, the two clusters with
the highest AGN fraction, Abell 89B and Abell 3125 have velocity dispersions of
$\sim 500 {\rm \, km s}^{-1}$, more typical of rich groups. In the {\it right}
panel of Figure~\ref{fig:frac_cl}, we compare the combined fractions of the two
low velocity dispersion clusters, $f_{\rm A}(M_R<-20;L_{X,B}>10^{41};\sigma<500)
= 0.100^{+0.062}_{-0.043}$, to the four higher velocity dispersion clusters,
$f_{\rm A}(M_R<-20;L_{X,B}>10^{41};\sigma>500) = 0.026^{+0.010}_{-0.008}$. Using
the binomial theorem to calculate the confidence intervals on the fractions
\citep[e.g.,][]{G1986}, we find that the probability the two above
fractions overlap is very small, $\sim 0.3\%$. However, this could overestimate
the significance of the result as there are fifteen different combinations of
two cluster groups we could make from our cluster sample. Therefore, we
conservatively estimate that the AGN fraction is higher in lower velocity
dispersion clusters at the $\sim95\%$ confidence level. A similar trend appears
in the Abell 901/902 supercluster \citet{GGO+2007}. There, X-ray AGN prefer to
be in regions with group-like environments (mainly based on galaxy density), as
compared to field-like or cluster-like environments. An anticorrelation between
optical AGN fraction and velocity dispersion
\citep{PB2006} and a correlation between radio AGN fraction and larger
environmental densities \citep{BKH+2005} have also been observed.
We do not expect that this result is sensitive to the varying radial coverage
between individual clusters. First, we found no radial dependence in the AGN
fraction at these luminosities. Second, the discrepancy between the fractions is
even more significant if we exclude Abell 644 and Abell 3128, the two clusters
with the least complete radial coverage.

Smaller X-ray AGN fractions than that found in our low velocity dispersion
clusters are measured in less dense environments. \citet{SMR+2007} only found
one X-ray AGN (out of 50 $M_R < -20$ galaxies) in a sample of eight $z \sim
0.06$ poor groups ($\sigma < 500 {\rm \, km s}^{-1}$), $f_{\rm
A}(M_R<-20;L_{X,B}>10^{41}) = 0.020^{+0.044}_{-0.017}$, where the majority of
these groups had smaller velocity dispersion than Abell 89B and Abell 3125. The
X-ray AGN fraction of early-type field galaxies in the Extended Chandra Deep
Field-South has also been measured \citep{LBA+2007}. They find $f_{\rm
A}(M_R<-20;L_{X,B}>10^{41}) = 0.066^{+0.034}_{-0.024}$ (B. Lehmer 2006, private
communication). We note that the X-ray AGN fraction for all $M_R < -20$ galaxies
drops by a factor of $\sim 2$ compared to the fraction for just early-type
galaxies in nearby clusters (T. Arnold et~al., in preparation). Since late-type
galaxies are more prevalent in the field than in clusters, one expects the field
X-ray AGN fraction for all galaxies to drop more rapidly compared to the $\sim
7\%$ measured for early-type galaxies.

We note that any additional obscuration associated with gas-rich galaxies will
be more prevalent where the fraction of late-type galaxies is higher. The effect
of missing AGN due to obscuration will be strongest in the field and weakest in
the highest velocity dispersion clusters. Thus, obscuration is unlikely to
explain the apparent prevalence of X-ray AGN in rich groups and poor clusters.

The likelihood of galaxy mergers increases with increasing galaxy density and
decreasing relative velocity. Compared to poor groups and the field, the
galaxy densities of rich groups are higher. Compared to galaxies in clusters,
the relative velocities of galaxies in rich groups are lower.
Thus, it is not surprising that AGN may form preferentially in group-like
environments. A larger sample of groups and clusters, particularly those with
velocity dispersions of poor clusters or rich groups at $\sigma \sim 500 {\rm \,
km/s}$, are needed to determine the preferred environment for AGN and use this
information to determine how they are fueled. Larger datasets of comparably
selected X-ray AGN in the field would also be valuable.

\section{Conclusions}
\label{sec:end}

To better understand the factors that may drive the evolution of AGN today, we
measure the AGN fraction in a new sample of nearby rich clusters, compare it to
more distant samples, and examine how it varies with environment. We present new
wide-field {\it Chandra} Observations of AGN in Abell 85 and Abell 754. 
Seventeen X-ray sources associated with galaxies in Abell 85, Abell 89B, and
Abell 754 are detected. Using $L_{X}/L_{K_s}$ relations we classify seven of
these galaxies as X-ray AGN with $L_{X,B} > 10^{41} {\rm \, erg \, s}^{-1}$. 
Only two of these X-ray AGN are classified as AGN based on their optical
spectra. Two of the X-ray galaxies not classified as X-ray AGN have been
previously spectroscopically identified as AGN. These sources are examples of
the additional AGN we expect are present in these clusters below our luminosity
threshold.

We add detections of X-ray AGN in three other $z \lesssim 0.08$ clusters, Abell
644, Abell 3125, and Abell 3128 to create a sample of sixteen X-ray AGN. We
find that $3.4^{+1.1}_{-0.8}\%$ of $M_R < -20$ galaxies host X-ray AGN with $L_{X,B} >
10^{41} {\rm \, erg \, s}^{-1}$. These results are consistent with the $\sim5\%$
fractions from \citetalias{MKK+2006,MMK2007}.
When compared to other samples \citep[e.g.,
\citetalias{MMK2007},][]{SJF+2007} at the same rest-frame X-ray luminosity and
absolute magnitude limits, there is no evidence for an evolving X-ray
AGN fraction for $z \lesssim 0.3$. 

Our three most important results are as follows:

\begin{itemize}
\item
{\bf X-ray AGN are detected more often in more luminous host galaxies.} This
observed trend of X-ray AGN fraction highlights the importance of using the same
X-ray luminosity and absolute magnitude cuts when comparing fractions.
This trend is due at least partially to a physical selection effect. Two
galaxies can have the same accretion rate relative to the Eddington limit, but
the galaxy with the more massive black hole will have a higher X-ray luminosity. 
More luminous host galaxies tend to have more massive black holes for a
combination of two reasons. First, the mass of a black hole scales with the mass
of a bulge component, and at a given bulge-to-disk ratio, a more luminous galaxy
will have a more massive bulge. Second, more luminous host galaxies are more
likely to be dominated by their spheroid.
Thus, it is not surprising that at a given X-ray luminosity one can detect less
efficient X-ray AGN in more luminous galaxies.
We require an understanding of
the distribution of accretion rates relative to the Eddington limit in X-ray AGN
to determine whether X-ray AGN are more likely to reside in more luminous host
galaxies for reasons beyond this physical selection effect.

\item
{\bf We do not find an excess fraction of X-ray AGN in the outskirts of
clusters.} The radial distribution of X-ray AGN appears to follow the same
distribution as cluster members without X-ray AGN. \citet{RE2005} also found a
relatively flat distribution of X-ray sources around massive, disturbed
clusters. The expectation from the major-merger or galaxy harassment pictures
for AGN fueling is that more AGN should be found in the outskirts of clusters. 
Acting against this expectation is our increased sensitivity to AGN in the most
luminous spheroids, as well as the significant fraction of early type galaxies
known to harbor low-luminosity AGN and LINERs. Thus, the tendency of the most
massive and early-type galaxies to lie in the centers of richest clusters could
dilute or even reverse AGN trends due to galaxy interactions. For instance,
\citet{RE2005} also found that massive, relaxed clusters had an excess
number of X-ray AGN in their central $0.5 {\rm \, Mpc}$ and near their virial
radius. The lack of any trend in our results could arise from the combination of
effects listed above, radial-averaging over the substructure in the cluster, or
that there is actually no preferred radial distribution for X-ray AGN.

\item
{\bf There is increasing evidence for higher AGN fractions in low velocity
dispersion clusters.} We find that the fraction of X-ray AGN is larger (at the
$\sim95\%$ confidence level) in lower velocity dispersion clusters or large
groups ($10.0^{+6.2}_{-4.3}\%$) than in richer clusters ($2.6^{+1.0}_{-0.8}\%$). 
Combined with results for the Abell 901/902 supercluster \citep{GGO+2007}, poor
groups \citep{SMR+2007}, and the field \citep{LBA+2007}, one can form a picture
where X-ray AGN in the local Universe are preferentially found in rich group
environments. If gas-rich mergers between galaxies are the principal driver of
AGN, then there should be more AGN in groups, where the galaxies tend to have
higher gas fractions and smaller relative velocities than in richer clusters,
but galaxy densities are higher than in the field. Because the most massive and
early type galaxies tend to lie in the richest clusters, and these galaxies
often harbor detectable LINERs that may be low-luminosity AGN, any increase in
AGN fraction due to galaxy-galaxy interactions could be even larger than what
we measure.
Finally, we note that obscuration of X-ray AGN in gas-rich galaxies is
not likely to be responsible for this trend, as more gas-rich galaxies are
expected in lower density environments.
\end{itemize}

These issues can be addressed through larger samples of X-ray AGN for clusters,
groups, and the field. For clusters, there are a couple of key properties that
need to be better sampled: an increased number of lower velocity dispersion
clusters and a larger number of disturbed and relaxed clusters observed out to
their virial radius. Such samples would directly address the radial distribution
of X-ray AGN and whether there is a preferred host environment. Specifically,
such data would be valuable to resolve the potential inconsistency between our
second and third highlighted results, which may be due to small number
statistics in the outskirts of clusters or our averaging over the substructure
in clusters. In addition, larger numbers of X-ray
AGN will constrain the underlying Eddington accretion efficiency distribution
and (host luminosity dependent?) AGN fraction by convolving these properties
with the galaxy luminosity function and comparing to the observed AGN fractions. 
These constraints in turn can be tested against AGN fueling mechanisms.

\acknowledgments

We thank Francesco Shankar for a useful discussion on accretion efficiency.
Support for this work was provided by the National Aeronautics and Space
Administration through {\it Chandra} awards GO4-5122A and GO6-7091X,
issued by the Chandra X-ray Observatory, which is operated by the Smithsonian
Astrophysical Observatory for and on behalf of NASA under contract NAS8-03060.
This research has made use of the NASA/IPAC Extragalactic Database (NED) which
is operated by the Jet Propulsion Laboratory, California Institute of
Technology, under contract with the National Aeronautics and Space
Administration.
The Second Palomar Observatory Sky Survey (POSS-II) was made by the California Institute of Technology with funds from the National Science Foundation, the National Geographic Society, the Sloan Foundation, the Samuel Oschin Foundation, and the Eastman Kodak Corporation.

{\it Facilities:} \facility{CXO (ACIS)}


\bibliography{ms}

\end{document}